\pdfoutput=1
\PassOptionsToPackage{dvipsnames}{xcolor}
\documentclass[acmsmall,screen]{acmart}

\usepackage[utf8]{inputenc} 
\usepackage{amsthm}
\usepackage{thmtools}
\usepackage{thm-restate}
\usepackage{parskip}
\usepackage{vwcol} 
\usepackage{array}
\usepackage{tabularx}
\usepackage{tikz} 
\usepackage{pifont}
\usepackage{makecell}
\usepackage{lineno}
\usepackage{stackengine}
\usepackage{diagbox}
\usepackage{multirow}
\usepackage{multicol}
\usepackage{mdframed}
\usepackage{listings}
\usepackage{subcaption}
\usepackage{paralist}
\usepackage{wrapfig}
\usepackage{pgfplots}
\usepackage[ruled,vlined,linesnumbered]{algorithm2e}
\DontPrintSemicolon
\SetKwProg{Fn}{Def}{:}{end}
\SetKw{Report}{report}
\SetKw{Break}{break}
\SetKwFunction{FCheckInternalConsistency}{CheckReadConsistency}
\SetKwFunction{FCheckRC}{CheckRC}
\SetKwFunction{FCheckRA}{CheckRA}
\SetKwFunction{FCheckRepeatableReads}{CheckRepeatableReads}
\SetKwFunction{FCheckCC}{CheckCC}
\SetKwFunction{FComputeHB}{ComputeHB}

\usepackage{ifthen}
\usepackage[capitalise]{cleveref}
\usepackage{todonotes}
\usepackage{pgfplotstable}
\usepackage{colortbl}
\usepackage[group-separator={,}, group-minimum-digits=4, text-series-to-math, propagate-math-font]{siunitx}
\usetikzlibrary {arrows.meta, graphs, positioning, quotes, backgrounds, shapes.misc, fit}
\usepgfplotslibrary{groupplots}
\tikzset{
    >=stealth,
    every edge/.append style={thick},
    invisible/.style={opacity=0},
    shift left/.style={
        decorate,decoration={simple line,raise=#1}
    },
    shift right/.style={
        decorate,decoration={simple line,raise=-1*#1}
    },
    transaction/.style={
        draw,
        fill=white,
        thick,
        rounded corners,
        text width=1.05cm,
    },
    session/.style={
        ->,
        very thick,
        >=latex,
    },
}
\newcommand{\tuple}[1]{\langle #1 \rangle}

\newcommand{\ToolName}{\operatorname{AWDIT}}
\newcommand{\cmark}{\ding{51}}%
\newcommand{\xmark}{\ding{55}}%
\newcommand{\cmarkgreen}{\textcolor{\darkgreen}{\cmark}}
\newcommand{\xmarkred}{\textcolor{\darkred}{\xmark}}
\makeatletter
\newcommand{\strequal}[2]{\pdf@strcmp{#1}{#2}==0}
\makeatother

\colorlet{SO}{black}
\colorlet{WR}{blue!70}
\colorlet{CO}{orange!90}
\colorlet{HB}{purple!90}
\def \darkgreen {black!30!green}
\def \darkred {black!20!red}

\newcommand{\po}{{\color{SO}\mathsf{po}}}
\newcommand{\so}{{\color{SO}\mathsf{so}}}
\newcommand{\co}{{\color{CO}\mathsf{co}}}
\renewcommand{\wr}{{\color{WR}\mathsf{wr}}}
\newcommand{\relop}[1]{\xrightarrow{#1}}
\newcommand{\relopT}[1]{\xrightarrow{#1}\hspace{-3pt}^{+} \hspace{3pt}}
\newcommand{\relopR}[1]{\xrightarrow{#1}\hspace{-3pt}^{?} \hspace{3pt}}

\newcommand{\indentityRel}[1]{[#1]}

\newcommand{\IsolationLevel}{\mathcal{I}}
\newcommand{\RC}{\mathsf{RC}}
\newcommand{\RA}{\mathsf{RA}}
\newcommand{\CC}{\mathsf{CC}}
\newcommand{\SIiso}{\mathsf{SI}}
\newcommand{\StrongerThan}{\sqsubseteq}

\newcommand{\numEvents}{n}
\newcommand{\numSessions}{k}
\newcommand{\numKeys}{\ell}
\newcommand{\KeySet}{\mathsf{Key}}
\newcommand{\ValSet}{\mathsf{Val}}
\newcommand{\OpIdSet}{\mathsf{OpId}}
\newcommand{\AllOps}{\mathsf{Op}}
\newcommand{\OpSet}{O}
\newcommand{\W}{\mathrm{W}}
\newcommand{\R}{\mathrm{R}}
\newcommand{\Read}[2][]{\R_{#1}(#2)}
\newcommand{\Write}[2][]{\W_{#1}(#2)}
\newcommand{\txnVar}{t}
\newcommand{\sessionVar}{s}
\newcommand{\readVar}{r}
\newcommand{\writeVar}{w}
\newcommand{\valVar}{v}
\newcommand{\opVar}{o}
\newcommand{\historyVar}{H}
\newcommand{\TxnSet}{T}
\newcommand{\TxnSetCommitted}{T_c}
\newcommand{\TxnSetAborted}{T_a}
\newcommand{\SessionsOf}[1]{\mathsf{sessions}(#1)}
\newcommand{\TxnsOfSession}[2]{{#1}|_{#2}}
\newcommand{\SessionOf}[1]{{#1}.\mathsf{sess}}
\newcommand{\ReadsOf}[1]{{#1}|_{\mathrm{R}}}
\newcommand{\WritesOf}[1]{{#1}|_{\mathrm{W}}}
\newcommand{\OpsOnKey}[2]{{#1}|_{#2}}
\newcommand{\ReadsOnKey}[2]{{#1}|_{\Read{#2}}}
\newcommand{\WritesOnKey}[2]{{#1}|_{\Write{#2}}}
\newcommand{\keyOf}[1]{#1.\mathsf{key}}
\newcommand{\valOf}[1]{#1.\mathsf{val}}
\newcommand{\txnOf}[1]{#1.\mathsf{txn}}
\newcommand{\keysWritten}[1]{\mathsf{KeysWt}({#1})}
\newcommand{\keysRead}[1]{\mathsf{KeysRd}({#1})}

\newcommand{\Paragraph}[1]{\smallskip\noindent{\bf #1}}

\newcommand{\earliestWriter}{\mathit{earliestWts}}
\newcommand{\readKeys}{\mathit{readKeys}}
\newcommand{\HB}{\mathit{HB}}
\newcommand{\Writes}[1]{\mathit{Writes}_{#1}}
\newcommand{\lastWrite}[1]{\mathit{lastWrite}_{#1}}
\newcommand{\lastWriter}{\mathit{lastWriter}}
\newcommand{\readTxns}{\mathit{readTxns}}
\newcommand{\firstTxnReads}{\mathit{firstTxnReads}}

\SetAlFnt{\small}
\IncMargin{1em}
\SetInd{0.28em}{0.28em}

\SetCommentSty{mycommfont}

\AtBeginDocument{%
    }

\setcopyright{cc}
\setcctype{by}
\acmJournal{PACMPL}
\acmYear{2025} \acmVolume{9} \acmNumber{PLDI} \acmArticle{236} \acmMonth{6} \acmPrice{}\acmDOI{10.1145/3729339}

\begin{document}

\title{AWDIT: An Optimal Weak Database Isolation Tester}

\author{Lasse M\o{}ldrup}
\orcid{0009-0005-9670-7039}
\affiliation{%
    \institution{Aarhus University}
    \city{Aarhus}
\country{Denmark}}
\email{moeldrup@cs.au.dk}

\author{Andreas Pavlogiannis}
\orcid{0000-0002-8943-0722}
\affiliation{%
    \institution{Aarhus University}
    \city{Aarhus}
\country{Denmark}}
\email{pavlogiannis@cs.au.dk}

\begin{abstract}
\emph{Database isolation} is a formal contract concerning the level of data consistency that a database provides to its clients.
In order to achieve low latency, high throughput, and partition tolerance, modern databases forgo strong transaction isolation for \emph{weak isolation} guarantees.
However, several production databases have been found to suffer from \emph{isolation bugs}, breaking their data-consistency contract.
\emph{Black-box testing} is a prominent technique for detecting isolation bugs, by checking whether histories of database transactions adhere to a prescribed isolation level.

In order to test databases on realistic workloads of large size, isolation testers must be as efficient as possible, a requirement that has initiated a study of the complexity of isolation testing.
Although testing strong isolation has been known to be NP-complete,
weak isolation levels were recently shown to be testable in polynomial time,
which has propelled the scalability of testing tools.
However, existing testers have a large polynomial complexity, restricting testing to workloads of only moderate size, which is not typical of large-scale databases.
\emph{How efficiently can we provably test weak database isolation?}

In this work we develop $\ToolName$, \emph{a highly-efficient and provably optimal tester for weak database isolation}.
Given a history $\historyVar$ of size $\numEvents$ and $\numSessions$ sessions, $\ToolName$ tests whether $\historyVar$ satisfies the most common weak isolation levels of Read Committed ($\RC$), Read Atomic ($\RA$), and Causal Consistency ($\CC$) in time $O(\numEvents^{3/2})$, $O(\numEvents^{3/2})$, and $O(\numEvents\cdot \numSessions)$, respectively, improving significantly over the state of the art.
Moreover, we prove that $\ToolName$ is essentially \emph{optimal}, in the sense that there is a lower bound of $n^{3/2}$, based on the combinatorial BMM hypothesis, for \emph{any} weak isolation level between $\RC$ and $\CC$.
Our experiments show that $\ToolName$ is significantly faster than existing, highly optimized testers; e.g., for the $\sim$20\% largest histories, AWDIT obtains an average speedup of $245\times$, $193\times$, and $62\times$ for $\RC$, $\RA$, and $\CC$, respectively, over the best baseline.

\end{abstract}


\maketitle

\section{Introduction}\label{SEC:INTRO}

Modern databases must handle enormous amounts of data, provide low latency, and be robust to network anomalies such as delays and partition faults.
To respond to such demands, databases typically forgo strong data consistency guarantees, such as any collection of concurrent database transactions admitting a serial view (i.e., being \emph{serializable}).
Instead, modern NewSQL and NoSQL databases support transactions that are weakly isolated, but ensure that the system remains available and efficient at all times (aka highly available transactions (HATs))~\cite{Bailis2013,Bailis2016,Akkoorath2016,Didona2018}.
The precise database guarantees of data integrity are specified as \emph{isolation levels} and have been subject to extensive formalization using various techniques such as axiomatic/graph-based~\cite{Terry1994a,Berenson1995,Adya2000,Bailis2016}, operational~\cite{Crooks2017}, and most recently, using an atomic visibility relation~\cite{Burckhardt2014,Cerone2015,Biswas2019}.
Common examples of weak isolation include 
Read Committed ($\RC$, the default level for most database transactions)~\cite{Bailis2013,Pavlo2017a}, 
Read Atomic ($\RA$)~\cite{Bailis2016,Cheng2021}, 
and (Transactional) Causal Consistency~\cite{{Akkoorath2016,Mehdi2017,Didona2018}} ($\CC$, available in, e.g., MongoDB~\cite{MongoDBCC}, Azure Cosmos~\cite{AzureCosmosCC} and Neo4j~\cite{Neo4j}).

Unfortunately, isolation levels can be tricky to understand and their implementation is error prone,
with isolation bugs being continuously discovered in production databases
~\cite{JepsenAnalyses,Kingsbury2020}.
The prevalence of isolation bugs has been targeted by database-testing techniques.
In particular, black-box testing is a popular approach, operating in two steps.
First, a client interacts with the database and records (logs) its history of interaction as a collection of transactions, each sending and receiving data to and from the database.
Second, an isolation tester analyzes the history and checks whether it adheres to the prescribed isolation level.
This process involves large histories spanning thousands to millions of transactions, in order to create a realistic load that is likely to expose an isolation anomaly.
As such, it has  spawned a research interest in isolation testers that are as efficient as possible, both in theory, by analyzing the computational complexity of database isolation testing, and in practice, by utilizing clever optimizations.

Testing strong isolation levels, such as Serializability and Snapshot Isolation, is known to be NP-complete (in the number of operations performed)~\cite{Papadimitriou1979a,Biswas2019}, leading most isolation testers to utilize SAT/SMT solvers~\cite{Tan2020,Zhang2023a,Huang2023b,Geng2024}.
On the other hand, weak isolation levels, such as Read Committed, Read Atomic, and Causal Consistency, have been shown to be checkable in polynomial time, allowing black-box testing algorithms of higher scalability, both in theory and in practice~\cite{Biswas2019}.
Elle is another popular database tester that runs in polynomial time and supports weak isolation levels~\cite{Kingsbury2020}, although its soundness is only guaranteed for certain types of transactions 
following ``list-append'' semantics.
The most recent development in this progression has been Plume~\cite{Liu2024a}, which appears to be the only algorithm stating an explicit polynomial complexity, which is of degree 6 (in particular, $O(\numEvents^3\cdot\numKeys^2\cdot \numSessions)$ for a history of $\numEvents$ transactions, $\numKeys$ keys and $\numSessions$ sessions).
Although other testers may possibly have a better complexity, Plume targets efficiency by utilizing efficient data structures including Vector Clocks~\cite{Friedemann1989} and Tree Clocks~\cite{Mathur2022}, and was shown to clearly outperform existing testers.

All these recent advances in database isolation testing highlight a demand for more performant testers, both in theory and in practice.
\emph{What is the precise complexity of testing weak database isolation?}
\emph{Are there provably optimal testing algorithms?}
We address this challenge in this work by developing $\ToolName$ (\underline{A} \underline{W}eak \underline{D}atabase \underline{I}solation \underline{T}ester):
\emph{a highly-efficient tester for weak database isolation} that is \emph{provably optimal} under standard assumptions.

\subsection{Motivating Example}\label{SUBSEC:MOTIVATING}

\begin{figure}
\def\scaleboxvalue{0.89}
\begin{subfigure}[b]{0.48\textwidth}
\newcommand{\sgap}{1.95}
    \newcommand{\tgap}{0.2}
    \newcommand{\slen}{3.22}
    \scalebox{\scaleboxvalue}{%
    \centering
    \begin{tikzpicture}[font=\small]
        \node[] (s1) at (0*\sgap, 0) {\small$\sessionVar_1$};
        \node[] (s2) at (1*\sgap, 0) {\small$\sessionVar_2$};
        \node[] (s3) at (2*\sgap, 0) {\small$\sessionVar_3$};
        \node[] (s4) at (3*\sgap, 0) {\small$\sessionVar_4$};
        \draw[session] (s1) -- ++(0,-\slen);
        \draw[session] (s2) -- ++(0,-\slen);
        \draw[session] (s3) -- ++(0,-\slen);
        \draw[session] (s4) -- ++(0,-\slen);

        \node[transaction, label=0:{\small$\txnVar_1$}, below = \tgap of s1] (t1) {
            $\Write{x, 1}$\\
            $\Write{y, 1}$
        };

        \node[transaction, label=0:{\small$\txnVar_2$}, below = \tgap of s2] (t2) {
            $\Write{x, 2}$
        };

        \node[transaction, label=0:{\small$\txnVar_3$}, below = \tgap of s3] (t3) {
            $\Write{x, 3}$
        };
        \node[transaction, label=0:{\small$\txnVar_4$}, below = \tgap of t3] (t4) {
            $\Write{z, 1}$\\
            $\Write{y, 2}$
        };

        \node[transaction, label=0:{\small$\txnVar_5$}, below = \tgap of s4] (t5) {
            $\Read{x, 1}$\\
            $\Read{x, 2}$\\
            $\Read{x, 3}$
        };
        \node[transaction, label=0:{\small$\txnVar_6$}, below = \tgap of t5] (t6) {
            $\Read{z, 1}$\\
            $\Read{y, 1}$
        };

        \draw (t1.north east) edge[draw=CO, ->, bend left, "$\co'$"] (t2.north west);
        
        \draw (t2.north east) edge[draw=CO, ->, bend left, "$\co'$"] (t3.north west);

        \draw (t4) edge[draw=CO, ->, bend left=10] node[below, pos=0.35]{$\co'$} (t1);

		\draw (t3) edge[draw=WR, ->, bend right=5] node[below]{$\wr_x$} (t5);
		\draw (t2) edge[draw=WR, ->, out=50, in=130, looseness=0.9, "$\wr_x$"] (t5);
		\draw (t1) edge[draw=WR, ->, out=55, in=125, looseness=0.9, "$\wr_x$"] (t5);

		\draw (t4) edge[draw=WR, ->, bend right=5] node[below]{$\wr_z$} (t6);
		\draw (t1) edge[draw=WR, ->, bend right=20] node[below]{$\wr_y$} (t6);

    \end{tikzpicture}
    }%
    \caption{$\RC$-inconsistent.}
    \label{subfig:intro_rc}
\end{subfigure}
\hfill
\begin{subfigure}[b]{0.48\textwidth}
    \newcommand{\sgap}{1.95}
    \newcommand{\tgap}{0.2}
    \newcommand{\slen}{3.22}
    
	 \scalebox{\scaleboxvalue}{%
	 \centering
    \begin{tikzpicture}[font=\small]
        \node[] (s1) at (0*\sgap, 0) {\small$\sessionVar_1$};
        \node[] (s2) at (1*\sgap, 0) {\small$\sessionVar_2$};
        \node[] (s3) at (2*\sgap, 0) {\small$\sessionVar_3$};
        \node[] (s4) at (3*\sgap, 0) {\small$\sessionVar_4$};
        \draw[session] (s1) -- ++(0,-\slen);
        \draw[session] (s2) -- ++(0,-\slen);
        \draw[session] (s3) -- ++(0,-\slen);
        \draw[session] (s4) -- ++(0,-\slen);

        \node[transaction, label=0:{\small$\txnVar_1$}, below = \tgap of s1] (t1) {
            $\Write{x, 1}$
        };
        \node[transaction, label=0:{\small$\txnVar_2$}, below = \tgap of t1] (t2) {
            $\Write{x, 2}$
        };
        \node[transaction, label=0:{\small$\txnVar_3$}, below = \tgap of t2] (t3) {
            $\Write{y, 1}$\\
            $\Read{z,2}$
        };

        \node[transaction, label=0:{\small$\txnVar_4$}, below = \tgap of s2] (t4) {
            $\Write{x, 3}$
        };
        \node[transaction, label=0:{\small$\txnVar_5$}, below = \tgap of t4] (t5) {
            $\Write{z, 1}$
        };

        \node[transaction, label=0:{\small$\txnVar_6$}, below = \tgap of s3] (t6) {
            $\Write{x, 4}$\\
            $\Read{z, 1}$\\
            $\Write{z, 2}$
        };

        \node[transaction, label=0:{\small$\txnVar_7$}, below = \tgap of s4] (t7) {
            $\Read{x, 3}$\\
            $\Read{y, 1}$
        };

        \draw (t6.north west) edge[draw=CO, ->, bend right, "$\co'$"'] (t4.north east);
        
        \draw (t2) edge[draw=CO, ->] node[below,anchor=north,xshift=5pt,yshift=1pt] {$\co'$} (t4);
        
        \draw (t5) edge[draw=WR, ->, bend right] node[below,anchor=north] {$\wr_z$} (t6);

        \draw (t4) edge[draw=WR, ->, out=45, "$\wr_x$"] (t7);

        \draw (t6) edge[draw=WR, ->, out=-105, in=15, looseness=0.7] node[below, anchor=north, xshift=-3pt] {$\wr_z$} (t3);

        \draw (t3) edge[draw=WR, ->, bend right, "$\wr_y$"'] (t7);
    \end{tikzpicture}
    }%
    \caption{$\CC$-inconsistent.}
    \label{subfig:intro_cc}
\end{subfigure}
\caption{\label{fig:intro}
An $\RC$-inconsistent history (\protect\subref{subfig:intro_rc}) and a $\CC$-inconsistent history (\protect\subref{subfig:intro_cc}).
$\ToolName$ infers a small set of partial commit edges $\co'$ that are sufficient to witness the inconsistency in each case and identify small witnesses by means of simple cycles.
Inferred $\co'$ edges that go along $\so\cup \wr$ are not shown explicitly.
}
\end{figure}
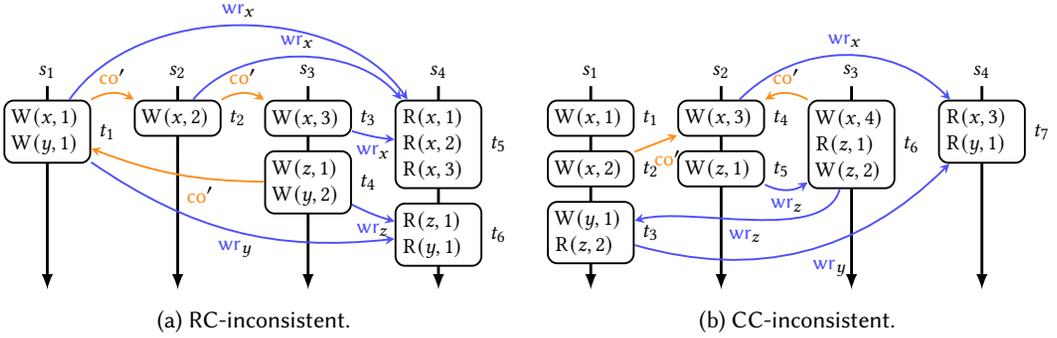

We illustrate violations of the Read Committed ($\RC$) and  Causal Consistency ($\CC$) isolation levels on two small histories in \cref{fig:intro}.
At a high level, the task of an isolation tester is to determine a \emph{commit order} $\co$ that is a total order on all transactions and satisfies certain properties, specific to the prescribed isolation level.
This $\co$ must also agree with the \emph{session order}, written $\so$, which totally orders the transactions of each session (shown in vertical black arrows), and the \emph{write-read order} $\wr$, which pairs transactions that common data is written by and read from (shown in blue).

Similarly to other testers, $\ToolName$ infers a \emph{partial relation} $\co'$ based on isolation-level-dependent inference rules.
Its key advantage lies in $\co'$ being small enough to be efficiently computable and yet sound and complete, in the sense that the history adheres to the isolation level iff $\co'$ is acyclic:~if not, a cycle in $\co'$ witnesses an isolation anomaly, whereas if yes, any total extension of $\co'$ serves as the commit order $\co$ that witnesses conformance to the isolation level (see \cref{def:saturated}).

Let us see why each of the two histories in \cref{fig:intro} violate their respective isolation levels and how $\ToolName$ determines this fact.

\Paragraph{Read Committed.}
$\RC$ states that (i)~only committed transactions can be read, and (ii)~a transaction $\txnVar$ cannot read a key $x$ from another transaction $\txnVar'$, if $\txnVar$ has previously observed (i.e., read a value from) a transaction that writes to $x$ and is $\co$-after $\txnVar'$.
Exploiting that $\co$ must be a total order, we can view this requirement as an inference rule:~if a transaction $\txnVar_3$ first observes some transaction $\txnVar_2$ that writes to $x$, and then $\txnVar_3$ reads $x$ from $\txnVar_1$, then (ii) implies (we infer) $\txnVar_2 \relop{\co} \txnVar_1$ for any $\RC$-consistent commit order $\co$ (see \cref{subfig:isolation-levels-rc} for a visual depiction).

Let us apply the above inference process to the history in \cref{subfig:intro_rc}, inferring the edges labeled $\co'$.
The fact that these edges form a cycle, when including that $\txnVar_3 \relop{\so} \txnVar_4$, then proves that \emph{no} total commit order exists, demonstrating that the history does not satisfy $\RC$.
Since $\txnVar_1 \relop{\wr_x} \txnVar_5$ (via $\Read{x,1}$) and later $\txnVar_2 \relop{\wr_x} \txnVar_5$ (via $\Read{x,2}$), we infer $\txnVar_1 \relop{\co} \txnVar_2$.
Similarly, since $\txnVar_2 \relop{\wr_x} \txnVar_5$ and later $\txnVar_3 \relop{\wr_x} \txnVar_5$, we infer $\txnVar_2 \relop{\co} \txnVar_3$.
Finally, since $\txnVar_4$ writes to $y$, $\txnVar_4 \relop{\wr_z} \txnVar_6$, and later $\txnVar_1 \relop{\wr_y} \txnVar_6$, we infer $\txnVar_4 \relop{\co} \txnVar_1$, completing the cycle.
$\ToolName$ constructs  a $\co'$ that contains exactly these three edges (as well as $\so$ and $\wr$ edges).
Notably, $\ToolName$ does not directly create some inferrable $\co'$ orderings, as long as they are present transitively (in $(\co')^+$), such as $\txnVar_1 \relop{\co'} \txnVar_4$ and $\txnVar_2 \relop{\co'} \txnVar_1$ in this example.
More importantly, it does not even need to check whether such transitive orderings are present.
Overall, $\ToolName$ spends only $O(\sqrt{\numEvents})$ time per transaction on average.

\Paragraph{Causal Consistency.}
Intuitively, $\CC$ states that transactions must obey causality:~if one transaction could have caused another, an observer should not observe the effect without also observing the cause.
Formally, a transaction $\txnVar_1$ is causally dependent on a transaction $\txnVar_2$, if there is a sequence of $\so$ and $\wr$ edges connecting $\txnVar_2$ to $\txnVar_1$, written succinctly as $t_2\relopT{\so\,\cup\,\wr}t_1$.
$\CC$ specifies that if a transaction $\txnVar$ reads a key $x$ from another transaction $\txnVar'$, then $\txnVar'$ must be the $\co$-latest among all transactions writing to $x$ that $\txnVar$ is causally dependent on.
We can also phrase this as an inference rule: if a transaction $\txnVar_3$ reads $x$ from another transaction $\txnVar_1$ and causally depends on a transaction $\txnVar_2$ that writes to $x$, we can infer $\txnVar_2 \relop{\co} \txnVar_1$ (see \cref{subfig:isolation-levels-cc} for a visual depiction).

Let us see how $\ToolName$ infers the $\co'$ in \cref{subfig:intro_cc} via the above inference rule.
Since $\txnVar_7$ reads $x$ from $\txnVar_4$, while $\txnVar_2$ and $\txnVar_6$ write to $x$, and  $\txnVar_2\relopT{\so\,\cup\,\wr} \txnVar_7$ and  $\txnVar_6\relopT{\so\,\cup\,\wr} \txnVar_7$, we have $\txnVar_2 \relop{\co'} \txnVar_4$ and $\txnVar_6\relop{\co'}\txnVar_4$.
The latter $\co'$ edge completes a cycle witnessing non-conformance to $\CC$.
Again, no further $\co'$ edges need to be inferred, with the guarantee that the existing ones represent all inferrable paths in the graph.
Finally, the inferred $\co'$ edges are computed in an efficient way that requires $O(\numSessions)$ time per transaction on average, where $\numSessions$ is the number of sessions.

\subsection{Our Contributions}\label{SUBSEC:CONTRIBUTIONS}

Here we state the main results of this paper, while we refer to the following sections for relevant definitions, algorithms, and lemmas.
All proofs are relegated to the Appendix.

\Paragraph{Upper bounds.}
First, we address the problem of testing the weak isolation levels Read Committed ($\RC$) and Read Atomic ($\RA$).
We consider histories of size $\numEvents$, measured as the number of read/write operations they contain.
We show that testing for $\RC$ and $\RA$ can be achieved in sub-quadratic time, which is much faster than existing isolation testers, as stated in the following theorem.

\begin{restatable}{theorem}{thmrarcupper}\label{thm:ra_rc_upper}
Given a history $\historyVar$ of size $\numEvents$, checking whether $\historyVar$ satisfies $\RC$ or $\RA$ can be decided in $O(\numEvents^{3/2})$ time.
\end{restatable}

We also remark that, when the size of each transaction is $O(1)$, the algorithms behind \cref{thm:ra_rc_upper} yield $O(\numEvents)$ running time.
Next, we turn our attention to the third common isolation level of Causal Consistency ($\CC$) and prove that it can be tested in quadratic time in general and in sub-quadratic time when the number of sessions is small, again, improving significantly over existing testers.

\begin{restatable}{theorem}{thmccupper}\label{thm:cc_upper}
Given a history $\historyVar$ of size $\numEvents$ and $\numSessions$ sessions, checking whether $\historyVar$ satisfies $\CC$ can be decided in $O(\numEvents\cdot \numSessions)$ time.
\end{restatable}

Normally, the number of sessions $\numSessions$ is significantly smaller than the number of operations $\numEvents$ of the history.
This stems from practical limitations of database deployment
and is also prevalent in database-testing benchmarks ~\cite{Kingsbury2020,Biswas2019,Liu2024a}.
In such cases, the bound of \cref{thm:cc_upper} becomes sub-quadratic and takes a linear form, when $\numSessions=O(1)$.

\Paragraph{Lower bounds.}
The above complexity improvements make it natural to ask:~\emph{Are further improvements possible? Is a linear bound possible for testing weak isolation?}
We now turn our attention to lower bounds, showing that \cref{thm:ra_rc_upper} and \cref{thm:cc_upper} are essentially (conditionally) optimal.

It is well known that Boolean Matrix Multiplication (BMM) can be computed in cubic time by the standard textbook algorithm.
The corresponding combinatorial BMM hypothesis states that there is no combinatorial algorithm achieving a truly sub-cubic bound for matrix multiplication~\cite{Williams19}.
Although the term ``combinatorial algorithm'' does not have a rigorous definition, it generally means an algorithm that does not rely on algebraic, fast matrix multiplication (FMM) techniques and must work irrespective of the structure over which the product is defined.

To state our first lower bound in its full generality, given two isolation levels $\IsolationLevel_1, \IsolationLevel_2$, we write $\IsolationLevel_1\StrongerThan \IsolationLevel_2$ to denote that $\IsolationLevel_1$ is stronger than $\IsolationLevel_2$, meaning that any history satisfying $\IsolationLevel_1$ also satisfies $\IsolationLevel_2$.

\begin{restatable}{theorem}{thmcclower}\label{thm:range_lower}
Consider any isolation level $\IsolationLevel$ with $\CC\StrongerThan \IsolationLevel\StrongerThan\RC$
and the problem of testing whether a history $\historyVar$ of size $\numEvents$ satisfies $\IsolationLevel$.
For any fixed $\epsilon>0$, there is
\begin{compactenum}
    \item\label{item:thm_range_lower_combinatorial} no combinatorial algorithm that runs in $O(\numEvents^{3/2-\epsilon})$ time, under the combinatorial BMM hypothesis, and
    \item\label{item:thm_range_lower_general} no algorithm that runs in $O(\numEvents^{\omega/2-\epsilon})$ time, where $\omega$ is the matrix multiplication exponent.
\end{compactenum}
\end{restatable}

\cref{thm:range_lower} is, perhaps, surprisingly general:~it states that the $n^{3/2}$ lower bound holds, not only for $\RA$, $\RC$, and $\CC$, but also for \emph{any} isolation level between them.
It further implies that, among combinatorial algorithms, our algorithms for $\RA$ and $\RC$ are \emph{optimal}, while our algorithm for $\CC$ may only be improved by a sub-linear factor $\sqrt{\numEvents}$.
Although this does not exclude faster isolation testers that use FMM, it is relevant for two reasons.
First, it has, thus far, been unclear whether FMM is useful in database testing.
Hence, our lower bound can be interpreted as ``unless we find a way to use FMM in isolation testing, the $O(\numEvents^{3/2})$  bound is likely tight''.
Second, although faster in theory, FMM is generally slow in practice because of large leading constants, and thus considered impractical.
Finally, \cref{item:thm_range_lower_general} of \cref{thm:range_lower} implies that a nearly linear-time tester (possibly relying on FMM) would be a major breakthrough, while a truly linear-time (i.e., $O(n)$) tester is \emph{impossible}~\cite{Coppersmith1982}.

Next, note that the lower bound of \cref{thm:range_lower} holds when the number of sessions $\numSessions$ is unbounded.
Zooming into each isolation level separately, we show that, in fact, $\RA$ retains its $\numEvents^{3/2}$ lower bound already with two sessions.

\begin{restatable}{theorem}{thmralower}\label{thm:ralower}
Consider the problem of testing whether a history $\historyVar$ of size $\numEvents$ and $2$ sessions satisfies $\RA$.
For any fixed $\epsilon>0$, there is
\begin{compactenum}
    \item no combinatorial algorithm that runs in $O(\numEvents^{3/2-\epsilon})$ time, under the combinatorial BMM hypothesis, and
    \item no algorithm that runs in $O(\numEvents^{\omega/2-\epsilon})$ time, where $\omega$ is the matrix multiplication exponent.
\end{compactenum}
\end{restatable}

Going one step further, we show that $\RC$ retains its $\numEvents^{3/2}$ lower bound even with just \emph{one session}.

\begin{restatable}{theorem}{thmrclower}\label{thm:rclower}
Consider the problem of testing whether a history $\historyVar$ of size $\numEvents$ and $1$ session satisfies $\RC$.
For any fixed $\epsilon>0$, there is
\begin{compactenum}
    \item no combinatorial algorithm that runs in $O(\numEvents^{3/2-\epsilon})$ time, under the combinatorial BMM hypothesis, and
    \item no algorithm that runs in $O(\numEvents^{\omega/2-\epsilon})$ time, where $\omega$ is the matrix multiplication exponent.
\end{compactenum}
\end{restatable}

\cref{thm:rclower} might be surprising, in the sense that analogous consistency problems for concurrent programs with a single thread are trivial (i.e., in linear time).
Finally, it is natural to ask how efficiently we can test  $\RA$ with only $\numSessions=1$ session.
Does it suffer, like $\RC$, the lower bound of $n^{3/2}$?
As the following theorem states, one-session histories are testable in linear time for $\RA$.

\begin{restatable}{theorem}{thmraonesessionlineartime}\label{thm:ra_one_session_linear_time}
Given a history $\historyVar$ of $\numEvents$ operations and $\numSessions=1$ session, checking whether $\historyVar$ satisfies $\RA$ can be decided in $O(\numEvents)$ time.
\end{restatable}

\Paragraph{In summary.}
Our results draw a fairly complete picture of the (fine-grained) complexity of weak database isolation testing.
In summary, for $\numSessions=1$ session, $\RA$ is testable in $O(\numEvents)$ time and is easier than $\RC$.
For $\numSessions\geq 2$, both $\RA$ and $\RC$ are testable in $O(\numEvents^{3/2})$ time.
Moreover, for \emph{any number of sessions}, our algorithms for testing $\RA$ and $\RC$ are (conditionally) \emph{optimal}.
Testing $\CC$ takes $O(\numEvents\cdot \numSessions)$ time and becomes super-linear only in the presence of many sessions,
whereas as the number of sessions grows, \emph{any} isolation level between $\RC$ and $\CC$ is unlikely to scale better than $\numEvents^{3/2}$.

\Paragraph{Implementation and experiments.}
We develop $\ToolName$, a prototype tool that implements our algorithms for testing weak isolation levels.
We evaluate the efficiency of $\ToolName$ on standard benchmarks
and compare its performance against all weak isolation testers from recent literature.
Our experiments reveal a clear advantage for $\ToolName$, which is always significantly faster and achieves speedups that exceed $1000\times$ in extreme cases, over all existing weak isolation testers.

\section{Preliminaries}\label{SEC:PRELIMINARIES}

We start with relevant definitions and notation regarding database transaction histories and weak isolation levels.
Our exposition mostly follows recent works~\cite{Biswas2019,Liu2024a}.

\subsection{Definitions}\label{SUBSEC:DEFINITIONS}

\Paragraph{Notation on relations.}
A (binary) relation $R$ over a set $X$ is a subset of $X \times X$. 
We write $x \relop{R} y$ to mean $\tuple{x, y} \in R$. 
The identity relation over $X$ is denoted by $\indentityRel{X} = \{\tuple{x,x} \mid x \in X\}$. 
The inverse of $R$ is $R^{-1}$. 
The reflexive closure and transitive closure of $R$ are $R^?$ and $R^+$, respectively, also written as $x \relopR{R} y$ and $x \relopT{R} y$. 
A relation $R$ over $X$ is \emph{irreflexive} if $\tuple{x, x} \notin R$ for all $x \in X$, and $R$ is \emph{acyclic} if $R^+$ is irreflexive.
For two relations $R_1, R_2$ over a common domain, we say that $R_1$ \emph{respects} $R_2$ (equivalently, $R_2$ respects $R_1$), if $R_1 \cup R_2$ is irreflexive.

\Paragraph{Databases.}
We consider transactional key-value databases over a set of keys $\KeySet = \{x,y,\dots\}$ and a set of values $\ValSet$.
Clients send operations to the database in the form of reads and writes. 
The set of possible operations for a set of keys $\KeySet$ and a set of values $\ValSet$ is denoted $\AllOps = \{ \Read[i]{x,\valVar}, \Write[i]{x,\valVar} \mid i \in \OpIdSet, x \in \KeySet, \valVar \in \ValSet \}$, where $\OpIdSet$ is a set of operation identifiers. 
When not relevant, we omit the operation identifier and simply write $\Read{x, \valVar}$ or $\Write{x, \valVar}$.
For brevity, we sometimes also refer to operations simply as $\readVar$, $\writeVar$, or $\opVar$, depending on if they are reads, writes, or arbitrary. 
In such cases, the key of an operation $\opVar$ is denoted by $\keyOf{\opVar}$, and its value by $\valOf{\opVar}$.

\Paragraph{Transactions.}
Client interactions with a database are grouped in transactions.
\begin{definition}\label{def:transaction}
A \emph{transaction} $\txnVar = \tuple{\OpSet, \po}$ is a set of operations $\OpSet \subseteq \AllOps$ and a \emph{program order} $\po$, which is a strict total order over $\OpSet$.
\end{definition}
For a transaction $\txnVar = \tuple{\OpSet, \po}$, the set of all read (resp. write) operations in $\txnVar$ is $\ReadsOf{\txnVar} = \{\Read{x,\valVar} \in \OpSet\}$ (resp. $\WritesOf{\txnVar} = \{\Write{x,\valVar} \in \OpSet\}$). 
This is naturally extended to sets of transactions $\TxnSet$, i.e., $\ReadsOf{\TxnSet} = \bigcup_{\txnVar \in \TxnSet} \ReadsOf{\txnVar}$ and $\WritesOf{\TxnSet} = \bigcup_{\txnVar \in \TxnSet} \WritesOf{\txnVar}$. 
The set of operations in $\txnVar$ acting on a key $x \in \KeySet$ is denoted by $\OpsOnKey{\txnVar}{x} = \{\opVar \in \OpSet \mid \keyOf{\opVar} = x \}$. 
The set of reads in $\txnVar$ reading a key $x \in \KeySet$ is denoted by $\ReadsOnKey{\txnVar}{x} = \ReadsOf{\txnVar} \cap \OpsOnKey{\txnVar}{x}$, and the set of writes in $\txnVar$ writing to $x$ is $\WritesOnKey{\txnVar}{x} = \WritesOf{\txnVar} \cap \OpsOnKey{\txnVar}{x}$. 
We also extend $\OpsOnKey{\TxnSet}{x}$, $\ReadsOnKey{\TxnSet}{x}$, and $\WritesOnKey{\TxnSet}{x}$ to sets of transactions $\TxnSet$ in the natural way. 
For $\opVar \in \OpSet$, we let $\txnOf{\opVar} = \txnVar$. 
The set of keys read (resp. written) by $\txnVar$ is denoted by $\keysRead{\txnVar}$ (resp. $\keysWritten{\txnVar}$). 
If $\txnVar$ contains a write to $x$, we say that $\txnVar$ writes $x$.

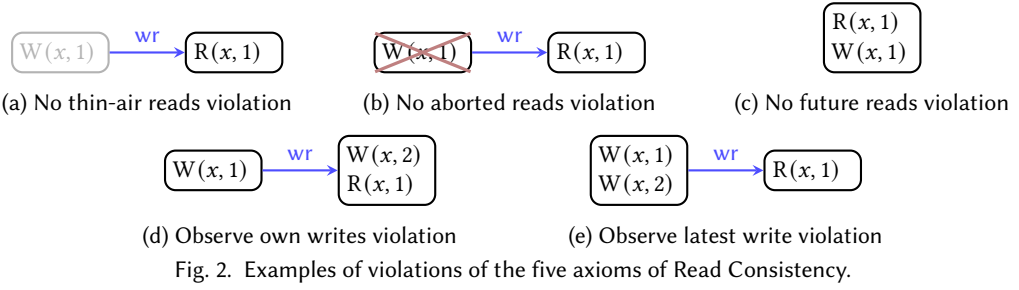
\Paragraph{Histories.}
At a high level, the collection of transactions between a database and its clients constitutes a \emph{history}. 
The \emph{session order}, written $\so$, over the transactions captures the total order of transactions executed in a single session and is thus a union of disjoint total orders.

In the setting of black-box database testing, the writes sent to the database are controlled by the tester. 
Since database implementations are normally data-independent~\cite{Wolper1986}, i.e., their behavior is independent of the concrete values written/read by the transactions, database testers use unique values on each write, because all isolation anomalies are preserved under this interaction scheme.
This implies that each read $\Read{x, v}$ observes the unique write $\Write{x,v}$ sent to the database in some (possibly remote) transaction.
Formally, the two events are related by the \emph{write-read} relation $\wr \subseteq \WritesOf{\TxnSet} \times \ReadsOf{\TxnSet}$, where $\TxnSet$ is the set of all transactions. 
We occasionally view $\wr$ as a relation on distinct transactions, i.e. $\txnVar_1 \relop{\wr} \txnVar_2$ iff
(i)~$\txnVar_1\neq \txnVar_2$, and
(ii)~$\writeVar \relop{\wr} \readVar$, where $\writeVar \in \WritesOf{\txnVar_1}$ and $\readVar \in \ReadsOf{\txnVar_2}$. 
We also write $\txnVar \relop{\wr} \readVar$ to denote that $\writeVar \relop{\wr} \readVar$ for some $\writeVar \in \WritesOf{\txnVar}$ and read $\readVar\not \in \txnVar$.
Finally, we project $\wr$ onto a specific key by writing $\wr_x = \wr \cap [\OpsOnKey{\TxnSet}{x}]$.
Transactions can either commit or abort; intuitively, an aborted transaction should not be visible to other transactions.

\begin{definition}\label{def:history}
    A \emph{history} $\historyVar = \tuple{\TxnSet, \so, \wr}$ is a set of transactions $\TxnSet = \TxnSetCommitted \uplus \TxnSetAborted$, a (strict partial) \emph{session order} $\so \subseteq \TxnSet \times \TxnSet$, and a write-read order $\wr \subseteq \WritesOf{\TxnSet} \times \ReadsOf{\TxnSet}$, where $\TxnSetCommitted$ is a set of committed transactions and $\TxnSetAborted$ is a set of aborted transactions. 
We require that $\wr^{-1}$ is a (partial) function.
\end{definition}

We let the set of sessions of $\historyVar$ be $\SessionsOf{\historyVar} = \{\sessionVar_1, \sessionVar_2, ..., \sessionVar_k\}$, and for a session $\sessionVar \in \SessionsOf{\historyVar}$, we let $\TxnsOfSession{\historyVar}{\sessionVar}$ be the \emph{committed} transactions of $\historyVar$ belonging to $\sessionVar$.
If $\txnVar \in \TxnsOfSession{\historyVar}{\sessionVar}$, we let $\SessionOf{\txnVar} = \sessionVar$.
The \emph{size} of $\historyVar$ is the total number of operations it contains.

\subsection{Weak Isolation Levels}\label{SUBSEC:WEAK_ISOLATION_LEVELS}

We follow the standard axiomatic approach of isolation specification using a commit order~\cite{Biswas2019}.
However, we are also interested in capturing more fine-grained transaction anomalies, which are assumed away in those axiomatic definitions.
For this purpose, we adapt some of the Transactional Anomalous Patterns (TAPs) proposed recently in~\cite{Liu2024a}.

\begin{figure}
    \centering
    \begin{subfigure}[b]{0.3\textwidth}
        \centering
        \begin{tikzpicture}[font=\small]
            \node[transaction, opacity=0.3] (t0) {$\Write{x, 1}$};
            \node[transaction, right = of t0] (t1) {$\Read{x, 1}$};
            \draw (t0) edge [draw=WR, "$\wr$", ->] (t1);
        \end{tikzpicture}
        \caption{No thin-air reads violation}
    \end{subfigure}
\hfill
    \begin{subfigure}[b]{0.3\textwidth}
        \centering
        \begin{tikzpicture}[font=\small]
            \node[transaction] (t0) {$\Write{x, 1}$};
            \node [
                fit=(t0),
                draw=red!50!black!50,
                very thick,
                inner sep=-\pgflinewidth,
                cross out,
            ] {};
            
            \node[transaction, right = of t0] (t1) {$\Read{x, 1}$};
            \draw (t0) edge [draw=WR, "$\wr$", ->] (t1);
        \end{tikzpicture}
        \caption{No aborted reads violation}
    \end{subfigure}
\hfill
    \begin{subfigure}[b]{0.3\textwidth}
        \centering
        \begin{tikzpicture}[font=\small]       
            \node[transaction, right = of t0] (t1) {
                $\Read{x, 1}$\\
                $\Write{x, 1}$
            };
        \end{tikzpicture}
        \caption{No future reads violation}
    \end{subfigure}
    \vspace{0.5em}\\
    \begin{subfigure}[b]{0.3\textwidth}
        \centering
        \begin{tikzpicture}[font=\small]
            \node[transaction,] (t0) {$\Write{x, 1}$};
            \node[transaction, right = of t0] (t1) {
                $\Write{x, 2}$\\
                $\Read{x, 1}$
            };
            \draw (t0) edge [draw=WR, "$\wr$", ->] (t1);
        \end{tikzpicture}
        \caption{Observe own writes violation}
    \end{subfigure}
\qquad\qquad
    \begin{subfigure}[b]{0.3\textwidth}
        \centering
        \begin{tikzpicture}[font=\small]
            \node[transaction,] (t0) {
                $\Write{x, 1}$\\
                $\Write{x, 2}$
            };
            \node[transaction, right = of t0] (t1) {
                $\Read{x, 1}$
            };
            \draw (t0) edge [draw=WR, "$\wr$", ->] (t1);
        \end{tikzpicture}
        \caption{Observe latest write violation}
    \end{subfigure}
    \caption{Examples of violations of the five axioms of Read Consistency.}
    \label{fig:internal-consistency}
\end{figure}
\Paragraph{Read Consistency.}
Read Consistency intuitively states that each read on $x$ observes either an earlier write on $x$ in its own transaction, or, if no such write exists, the last write on $x$ of a committed transaction\footnote{This is equivalent to disallowing G1a and G1b in~\cite{Adya2000}, in addition to their basic assumptions on histories.}.
Formally, this is stated as five basic axioms (illustrated in \cref{fig:internal-consistency} as TAPs).

\begin{definition}[Read Consistency]\label{def:internal-consistency}
A history $\historyVar = \tuple{\TxnSet, \so, \wr}$ satisfies \emph{Read Consistency} if the following conditions hold.
\begin{compactenum}
    \item[(a)] No thin-air reads: $\forall \readVar \in \ReadsOf{\TxnSetCommitted}, \exists \writeVar \in \WritesOf{\TxnSet}\colon \writeVar \relop{\wr} \readVar$.
    \item[(b)] No aborted reads: $\forall \readVar \in \ReadsOf{\TxnSetCommitted}, \forall \writeVar \in \WritesOf{\TxnSet}\colon \writeVar \relop{\wr} \readVar \implies \writeVar \notin \WritesOf{\TxnSetAborted}$.
    \item[(c)] No future reads: $\forall \readVar \in \ReadsOf{\TxnSetCommitted}, \forall \writeVar \in \WritesOf{\TxnSet}\colon \writeVar \relop{\wr} \readVar \implies \lnot(\readVar \relop{\po} \writeVar)$.
    \item[(d)] Observe own writes:
    $
        \forall \readVar \in \ReadsOf{\TxnSetCommitted}, \forall \writeVar \in \WritesOf{\TxnSet}\colon \writeVar \relop{\wr} \readVar \land \txnOf{\writeVar} \neq \txnOf{\readVar}\implies \not\exists \writeVar' \in \WritesOnKey{\TxnSet}{\keyOf{\readVar}} \colon \writeVar' \relop{\po} \readVar
    $.
    \item[(e)] Observe latest write:
    $
        \forall \readVar \in \ReadsOf{\TxnSetCommitted}, \forall \writeVar, \writeVar' \in \WritesOnKey{\TxnSet}{\keyOf{\readVar}} \colon \writeVar \relop{\wr} \readVar \land\writeVar \relop{\po} \writeVar' \implies \readVar \relop{\po} \writeVar'
    $.
\end{compactenum}
\end{definition}

We are now ready to define the three main weak isolation levels of Read Committed, Read Atomic, and Causal Consistency.
Each level requires Read Consistency, as well as an additional axiom involving a \emph{commit order} $\co$, which is a strict total order over all committed transactions that respects $\so\cup \wr$ and also satisfies a predicate specific to the isolation level at hand.

\Paragraph{Read Committed ($\RC$).}
The Read Committed\footnote{Some literature~\cite{Crooks2017} interprets $\RC$ as proscribing G1 from~\cite{Adya2000}, which is the weaker requirement of Read Consistency plus acyclicity of $\so\cup \wr$.
This is easily checkable in $O(\numEvents)$ time, for a history of size $\numEvents$.} isolation level formalizes the intuition that the database can only read from committed transactions, and also adheres to a monotonicity requirement:~a transaction $\txnVar$ is not allowed to read a key $x$ from another transaction $\txnVar'$, if it has previously observed (i.e., read a value from) a transaction that writes to $x$ and is $\co$-later than $\txnVar'$.

\begin{definition}[Read Committed]\label{def:rc}
    A history $\historyVar = \tuple{\TxnSet, \so, \wr}$ satisfies \emph{Read Committed} ($\RC$), if it is Read Consistent, and there is a strict total \emph{commit order} $\co$ over $\TxnSetCommitted$ respecting $\so \cup \wr$, such that the following holds (see \cref{subfig:isolation-levels-rc} for a pictorial depiction).
    \begin{align*}
        \forall x \in \KeySet, \forall &\txnVar_1, \txnVar_2 \in \TxnSetCommitted, \forall \readVar, \readVar_x \in \ReadsOf{\TxnSetCommitted}\colon\\
			& \txnVar_1 \neq \txnVar_2\ \land \txnVar_1 \relop{\wr_x} \readVar_x \land \text{$\txnVar_2$ writes $x$}\ \land 
            \txnVar_2 \relop{\wr} \readVar \relop{\po} \readVar_x 
            \quad \implies\quad  \txnVar_2 \relop{\co} \txnVar_1\ .
    \end{align*}
\end{definition}

\begin{example}
The history in \cref{subfig:isolation-level-examples-none} does not satisfy $\RC$.
In particular, $\txnVar_1\relop{\so}\txnVar_2$ forces that $\txnVar_1 \relop{\co} \txnVar_2$.
Hence, the second read of $x$ in $\txnVar_3$ should read $\txnVar_2$ instead of $\txnVar_1$.
The history in \cref{subfig:isolation-level-examples-rc}, on the other hand, satisfies $\RC$. 
Even though $\txnVar_3$ only observes the latter of the writes in $\txnVar_2$, $\txnVar_1$ is observed first, so there is no violation.
\end{example}

\begin{figure}
    \centering
    \begin{subfigure}[b]{0.3\textwidth}
        \begin{tikzpicture}[font=\small]
            \graph [math nodes, grow right=6em, branch down=4em]  {
                t_2["writes $x$"] ->[draw=WR, "$\wr$"] r[as=$\readVar$],
                {[nodes={xshift=2em}] t_1 ->[draw=WR, "$\wr_x$"] r_x[as=$\readVar_x$]},
                t_2 ->[draw=CO, "$\co$"] t_1,
                r ->["$\po$"] r_x
            };
            \draw[dashed] (r.center) ++(-0.37,0.25) -- ++(0.93,0) -- node [midway, label=0:$t_3$] {} ++(0.94,-1.9) -- ++(-0.93,0) -- cycle;
        \end{tikzpicture}
        \caption{Read Committed}
        \label{subfig:isolation-levels-rc}
    \end{subfigure}
    \begin{subfigure}[b]{0.3\textwidth}
        \begin{tikzpicture}[font=\small]
            \graph [math nodes, grow right=6em, branch down=4em]  {
                t_2["writes $x$"],
                {[nodes={xshift=2em}] t_1 ->[draw=WR, "$\wr_x$"] t_3},
                t_2 ->[draw=CO, "$\co$"] t_1,
                t_2 ->[bend left, "$\so \cup \wr$"] t_3
            };
        \end{tikzpicture}
        \caption{Read Atomic}
        \label{subfig:isolation-levels-ra}
    \end{subfigure}
    \begin{subfigure}[b]{0.3\textwidth}
        \begin{tikzpicture}[font=\small]
            \graph [math nodes, grow right=6em, branch down=4em]  {
                t_2["writes $x$"],
                {[nodes={xshift=2em}] t_1 ->[draw=WR, "$\wr_x$"] t_3},
                t_2 ->[draw=CO, "$\co$"] t_1,
                t_2 ->[bend left, dashed, "$(\so \cup \wr)^+$"] t_3
            };
        \end{tikzpicture}
        \caption{Causal Consistency}
        \label{subfig:isolation-levels-cc}
    \end{subfigure}
    \caption{The axioms of Read Committed (\protect\subref{subfig:isolation-levels-rc}), Read Atomic (\protect\subref{subfig:isolation-levels-ra}), and Causal Consistency (\protect\subref{subfig:isolation-levels-cc}).
In each case, the $\co$ ordering is required when the other orderings hold.
}
    \label{fig:isolation-levels}
\end{figure}
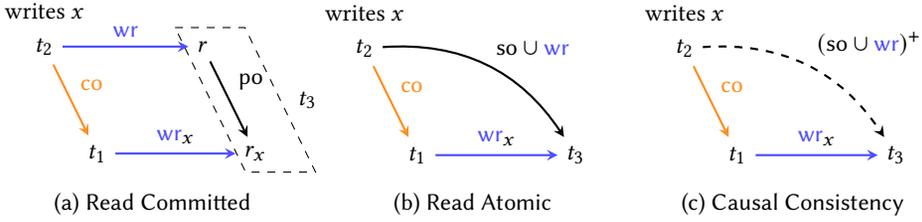
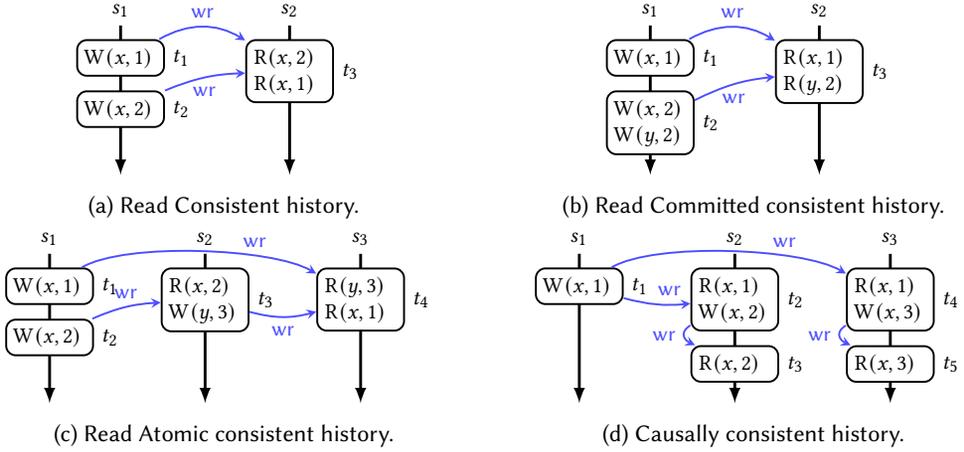
\begin{figure}
    \centering
	\def\scaleboxvalue{0.9}
    \begin{subfigure}[b]{0.45\textwidth}
        \centering
        \newcommand{\sgap}{2.5}
        \newcommand{\tgap}{0.2}
        \newcommand{\slen}{2.4}
		\scalebox{\scaleboxvalue}{%
        \begin{tikzpicture}[font=\small]
            \node[] (s1) at (0*\sgap, 0) {\small$\sessionVar_1$};
            \node[] (s2) at (1*\sgap, 0) {\small$\sessionVar_2$};
            \draw[session] (s1) -- ++(0,-\slen);
            \draw[session] (s2) -- ++(0,-\slen);

            \node[transaction, label=0:{\small$\txnVar_1$}, below = \tgap of s1] (t1) {
                $\Write{x,1}$
            };
            \node[transaction, label=0:{\small$\txnVar_2$}, below = \tgap of t1] (t2) {
                $\Write{x,2}$
            };

            \node[transaction, label=0:{\small$\txnVar_3$}, below = \tgap of s2] (t3) {
                $\Read{x,2}$\\
                $\Read{x,1}$
            };
			\draw (t1) edge [draw=WR, bend left=30, ->] node[above]  {$\wr$} (t3);
			\draw (t2) edge [draw=WR, bend left=10, ->] node[below]  {$\wr$} (t3);	
        \end{tikzpicture}
		}%
        \caption{Read Consistent history.}
        \label{subfig:isolation-level-examples-none}
    \end{subfigure}
	\qquad
    \begin{subfigure}[b]{0.45\textwidth}
        \centering
        \newcommand{\sgap}{2.5}
        \newcommand{\tgap}{0.2}
        \newcommand{\slen}{2.4}
		\scalebox{\scaleboxvalue}{%
        \begin{tikzpicture}[font=\small]
            \node[] (s1) at (0*\sgap, 0) {\small$\sessionVar_1$};
            \node[] (s2) at (1*\sgap, 0) {\small$\sessionVar_2$};
            \draw[session] (s1) -- ++(0,-\slen);
            \draw[session] (s2) -- ++(0,-\slen);

            \node[transaction, label=0:{\small$\txnVar_1$}, below = \tgap of s1] (t1) {
                $\Write{x,1}$
            };
            \node[transaction, label=0:{\small$\txnVar_2$}, below = \tgap of t1] (t2) {
                $\Write{x,2}$\\
                $\Write{y,2}$
            };

            \node[transaction, label=0:{\small$\txnVar_3$}, below = \tgap of s2] (t3) {
                $\Read{x,1}$\\
                $\Read{y,2}$
            };
			\draw (t1) edge [draw=WR, bend left=30, ->] node[above]  {$\wr$} (t3);
			\draw (t2) edge [draw=WR, bend left=10, ->] node[below]  {$\wr$} (t3);	
        \end{tikzpicture}
		}%
        \caption{Read Committed consistent history.}
        \label{subfig:isolation-level-examples-rc}
    \end{subfigure}\\
    \begin{subfigure}[b]{0.45\textwidth}
        \centering
        \newcommand{\sgap}{2.3}
        \newcommand{\tgap}{0.2}
        \newcommand{\slen}{2.4}
		\scalebox{\scaleboxvalue}{%
        \begin{tikzpicture}[font=\small]
            \node[] (s1) at (0*\sgap, 0) {\small$\sessionVar_1$};
            \node[] (s2) at (1*\sgap, 0) {\small$\sessionVar_2$};
            \node[] (s3) at (2*\sgap, 0) {\small$\sessionVar_3$};
            \draw[session] (s1) -- ++(0,-\slen);
            \draw[session] (s2) -- ++(0,-\slen);
            \draw[session] (s3) -- ++(0,-\slen);

            \node[transaction, label=0:{\small$\txnVar_1$}, below = \tgap of s1] (t1) {
                $\Write{x, 1}$
            };
            \node[transaction, label=0:{\small$\txnVar_2$}, below = \tgap of t1] (t2) {
                $\Write{x, 2}$
            };

            \node[transaction, label=0:{\small$\txnVar_3$}, below = \tgap of s2] (t3) {
                $\Read{x, 2}$\\
                $\Write{y, 3}$
            };

            \node[transaction, label=0:{\small$\txnVar_4$}, below = \tgap of s3] (t4) {
                $\Read{y, 3}$\\
                $\Read{x, 1}$
            };
			\draw (t1) edge [draw=WR, out=30, in=150, looseness=0.55, ->] node[above, pos=0.7]  {$\wr$} (t4);
			\draw (t2) edge [draw=WR, bend left=10, ->] node[above]  {$\wr$} (t3);
			\draw (t3) edge [draw=WR, bend right=15, ->] node[below]  {$\wr$} (t4);	
        \end{tikzpicture}
		}%
        \caption{Read Atomic consistent history.}
        \label{subfig:isolation-level-examples-ra}
    \end{subfigure}
		\qquad
    \begin{subfigure}[b]{0.45\textwidth}
        \centering
        \newcommand{\sgap}{2.3}
        \newcommand{\tgap}{0.2}
        \newcommand{\slen}{2.4}
		\scalebox{\scaleboxvalue}{%
        \begin{tikzpicture}[font=\small]
            \node[] (s1) at (0*\sgap, 0) {\small$\sessionVar_1$};
            \node[] (s2) at (1*\sgap, 0) {\small$\sessionVar_2$};
            \node[] (s3) at (2*\sgap, 0) {\small$\sessionVar_3$};
            \draw[session] (s1) -- ++(0,-\slen);
            \draw[session] (s2) -- ++(0,-\slen);
            \draw[session] (s3) -- ++(0,-\slen);

            \node[transaction, label=0:{\small$\txnVar_1$}, below = \tgap of s1] (t1) {
                $\Write{x, 1}$
            };

            \node[transaction, label=0:{\small$\txnVar_2$}, below = \tgap of s2] (t2) {
                $\Read{x, 1}$\\
                $\Write{x, 2}$
            };

            \node[transaction, label=0:{\small$\txnVar_3$}, below = \tgap of t2] (t3) {
                $\Read{x, 2}$
            };

            \node[transaction, label=0:{\small$\txnVar_4$}, below = \tgap of s3] (t4) {
                $\Read{x, 1}$\\
                $\Write{x, 3}$
            };

            \node[transaction, label=0:{\small$\txnVar_5$}, below = \tgap of t4] (t5) {
                $\Read{x, 3}$
            };

			\draw (t1) edge [draw=WR, out=30, in=150, looseness=0.55, ->] node[above, pos=0.7]  {$\wr$} (t4);
			\draw (t1) edge [draw=WR, bend right=10, ->] node[above, pos=0.7]  {$\wr$} (t2);
			\draw (t2) edge [draw=WR, out=-150, in=155, looseness=1.6, ->] node[left]  {$\wr$} (t3);
			\draw (t4) edge [draw=WR, out=-150, in=155, looseness=1.6, ->] node[left]  {$\wr$} (t5);
        \end{tikzpicture}
		}%
        \caption{Causally consistent history.}
        \label{subfig:isolation-level-examples-cc}
    \end{subfigure}
    \caption{Examples of consistent histories that violate consistency of stronger isolation levels.}
    \label{fig:isolation-level-examples}
\end{figure}

\Paragraph{Read Atomic ($\RA$).}
The Read Atomic isolation level formalizes the intuition that transactions should be atomic, in the sense that either all or none of the effects of a transaction can be observed.

\begin{definition}[Read Atomic]\label{def:ra}
    A history $\historyVar = \tuple{\TxnSet, \so, \wr}$ satisfies \emph{Read Atomic} ($\RA$), if it is Read Consistent, and there is a strict total \emph{commit order} $\co$ over $\TxnSetCommitted$ respecting $\so \cup \wr$, such that the following holds (see \cref{subfig:isolation-levels-ra} for a pictorial depiction).
    \begin{align*}
        \forall x \in \KeySet, \forall &\txnVar_1, \txnVar_2, \txnVar_3 \in \TxnSetCommitted\colon\quad
		 \txnVar_1 \neq \txnVar_2\ \land \txnVar_1 \relop{\wr_x} \txnVar_3 \land \text{$\txnVar_2$ writes $x$}\ \land 
            \txnVar_2 \relop{\so\,\cup\,\wr} \txnVar_3
            \quad\implies\quad \txnVar_2 \relop{\co} \txnVar_1\ .
    \end{align*}
\end{definition}

\begin{example}
Consider again the history in \cref{subfig:isolation-level-examples-rc}.
Transaction $\txnVar_3$ reads $y$ from $\txnVar_2$, but does not read its write to $x$, instead reading the older version written by $\txnVar_1$.
Hence, $\txnVar_3$ observes some, but not all, effects of $\txnVar_2$, violating $\RA$.
The history in \cref{subfig:isolation-level-examples-ra}, on the other hand, satisfies $\RA$.
Even though $\txnVar_4$ displays weak behavior by reading from $\txnVar_1$ instead of $\txnVar_2$,
it observes all effects of the transactions that it directly reads from.
\end{example}

\Paragraph{Causal Consistency ($\CC$).}
Causal Consistency\footnote{Sometimes also called Transactional Causal Consistency~\cite{Akkoorath2016, Liu2024a}.} specifies that reads must respect causal relationships between transactions:~intuitively, if a transaction $t$ reads a key $x$ from another transaction $t'$, then $t'$ must be the $\co$-latest among all transactions that $t$ is causally dependent on, and write to $x$.
The notion of causality is formalized via the \emph{happens-before} relation, dictating that transaction $\txnVar_1$ happens before transaction $\txnVar_2$, if $\txnVar_1 \relopT{\so\,\cup\,\wr} \txnVar_2$.

\begin{definition}[Causal Consistency]\label{def:cc}
    A history $\historyVar = \tuple{\TxnSet, \so, \wr}$ satisfies \emph{Causal Consistency} ($\CC$), if it is Read Consistent, and there is a strict total \emph{commit order} $\co$ over $\TxnSetCommitted$ respecting $\so \cup \wr$, such that the following holds (see \cref{subfig:isolation-levels-cc} for a pictorial depiction).
    \begin{align*}
        \forall x \in \KeySet, \forall &\txnVar_1, \txnVar_2, \txnVar_3 \in \TxnSetCommitted\colon \quad
			 \txnVar_1 \neq \txnVar_2\ \land
            \txnVar_1 \relop{\wr_x} \txnVar_3 \land \text{$\txnVar_2$ writes $x$}\ \land 
            \txnVar_2 \relopT{\so\,\cup\,\wr} \txnVar_3 
            \quad\implies\quad \txnVar_2 \relop{\co} \txnVar_1\ .
    \end{align*}
\end{definition}

\begin{example}
Consider again the history in \cref{subfig:isolation-level-examples-ra}, which does not satisfy $\CC$.
Transaction $\txnVar_4$ observes $\txnVar_2$ through its read on $y$, and it should therefore not observe $\txnVar_1$, which happens before $\txnVar_2$.
The history in \cref{subfig:isolation-level-examples-cc}, on the other hand, satisfies $\CC$.
Note that there is still weak behavior, however, as both $\txnVar_2$ and $\txnVar_4$ read a value of 1 on $x$ and then overwrite it, making the history non-serializable.
\end{example}

\Paragraph{Comparison of isolation levels.}
Given two isolation levels $\IsolationLevel_1$, $\IsolationLevel_2$, we say that $\IsolationLevel_1$ is \emph{stronger than} $\IsolationLevel_2$, denoted by $\IsolationLevel_1\StrongerThan\IsolationLevel_2$, if any history that satisfies $\IsolationLevel_1$ also satisfies $\IsolationLevel_2$.

\Paragraph{The consistency problem.}
The primary task of a back-box isolation tester is consistency checking:~given an isolation level $\IsolationLevel\in \{\RC, \RA, \CC\}$ and history $\historyVar$, decide whether $\historyVar$ satisfies $\IsolationLevel$.

\section{Weak Isolation Algorithms}\label{SEC:ALGORITHMS}

We now present algorithms for checking consistency under $\RC$, $\RA$, and $\CC$, towards \cref{thm:ra_rc_upper} and \cref{thm:cc_upper}.
Each algorithm starts by checking the axioms of Read Consistency (\cref{fig:internal-consistency}).
Given a history of size $\numEvents$, this check can easily be carried out in $O(\numEvents)$ time.
The precise algorithm for this task is delegated to \cref{SEC:APP_ALGORITHMS} (\cref{alg:internal-consistency}).

Each axiom of $\RC$, $\RA$, and $\CC$ requires the existence of a commit order $\co$ satisfying certain properties (\cref{fig:isolation-levels}).
For an input history $\historyVar$, the respective algorithm builds a \emph{partial} commit relation $\co'$ that holds \emph{necessary} orderings, in the sense that \emph{any} $\co$ witnessing the consistency of $\historyVar$ satisfies $\co'\subseteq \co$.
This implies that, if $\co'$ is cyclic, then $\historyVar$ is inconsistent.
Moreover, at the end of the algorithm's execution, the orderings in $\co'$ are also \emph{sufficient}, in the sense that if $\co'$ is acyclic, any linearization of $\co'$ serves as the total commit order $\co$ witnessing the consistency of $\historyVar$.
The key property of $\co'$ is that it is \emph{saturated} and \emph{minimal}, as defined below.

\begin{definition}[Saturated and minimal commit relations]\label{def:saturated}
Given an isolation level $\IsolationLevel \in \{\RC, \RA, \CC\}$ and a history $\historyVar = \tuple{\TxnSet, \so, \wr}$, a (partial) commit relation $\co'$ is \emph{saturated} for $\IsolationLevel$ if 
(i)~$\so \cup \wr \subseteq \co'$, and 
(ii)~if the premise in \cref{fig:isolation-levels} holds for $\IsolationLevel$, for transactions $\txnVar_1$, $\txnVar_2$, and $\txnVar_3$ (i.e., the respective figure without the $\co$ edge), then $\txnVar_2 \relopT{\co'} \txnVar_1$.
Moreover, $\co'$ is \emph{minimal} for $\IsolationLevel$ if, for any transactions $\txnVar_1$ and $\txnVar_2$ with $\txnVar_2 \relop{\co'} \txnVar_1$, either $\txnVar_2 \relop{\so\,\cup\,\wr} \txnVar_1$ or \cref{fig:isolation-levels} requires $\txnVar_2 \relop{\co} \txnVar_1$ for $\IsolationLevel$ (possibly both).
\end{definition}
We note that saturated relations for consistency exist in the literature (e.g.,~\cite{Biswas2019}), but our $\co'$ has an advantage due to its minimality, which allows for more efficient algorithms.
The correctness of the presented algorithms is based on the fact that saturated and minimal commit relations exactly characterize the consistency of $\historyVar$, as stated in the following lemma.
\begin{restatable}{lemma}{lemsaturated}\label{lem:saturated}
Given an isolation level $\IsolationLevel \in \{\RC, \RA, \CC\}$, a history $\historyVar$, and a minimal saturated commit relation $\co'$, $\historyVar$ satisfies $\IsolationLevel$ iff $\historyVar$ satisfies Read Consistency and $\co'$ is acyclic.
\end{restatable}

\subsection{Read Committed}\label{SUBSEC:ALGORITHMS_RC}

\begin{algorithm}
    \caption{Read Committed}
    \label{alg:rc}

    \Fn{\FCheckRC{$\historyVar = \tuple{\TxnSet, \so, \wr}$}}{
		\tcp{Algorithm \ref{alg:internal-consistency}}
        \FCheckInternalConsistency{$\historyVar$}\\  \label{line:rc_internal_consistency}
        $\co' \gets \so \cup \wr$ \; \label{line:rc_co_init}
        \For{$\txnVar_3 = \tuple{\OpSet, \po} \in \TxnSetCommitted$}{ \label{line:rc_t3_loop}
            $\readTxns \gets \emptyset$ ;
            $\firstTxnReads \gets \emptyset$ \;
            \For{$\readVar \in \ReadsOf{\txnVar_3}$ in $\po$ order}{ \label{line:rc_r_loop}
                Let $\txnVar_2$ be such that $\txnVar_2 \relop{\wr} \readVar$ \;
                \If{$\txnVar_2 \notin \readTxns$}{
                    $\readTxns \gets \readTxns \cup \{ \txnVar_2 \}$ \; \label{line:rc_read_txns_update}
                    $\firstTxnReads \gets \firstTxnReads \cup \{ \readVar \}$ \; \label{line:rc_first_txn_reads_update}
                }
            }
            $\earliestWriter \gets \lambda x. \tuple{\bot, \bot}$ ; \label{line:rc_earliest_writer_def}
            $\readKeys \gets \emptyset$ \;
            \For{$\Read{y, \valVar} \in \ReadsOf{\txnVar_3}$ in reverse $\po$ order}{ \label{line:rc_read_loop}
                Let $\txnVar_2$ be such that $\txnVar_2 \relop{\wr} \Read{y, \valVar}$ \;
                \If{$\Read{y, \valVar} \in \firstTxnReads$}{ \label{line:rc_first_txn_check}
                    \tcp{Loop over the smaller set}
                    \For{$x \in \keysWritten{\txnVar_2} \cap \readKeys$}{ \label{line:rc_x_loop}
                        $\txnVar_1 \gets \earliestWriter[x][1]$ \; \label{line:rc_t1_def_fst}
                        \lIf{$\txnVar_1 = \txnVar_2$}{\label{line:rc_t1_t2_check}
                            $\txnVar_1 \gets \earliestWriter[x][0]$ \label{line:rc_t1_def_snd}
                        }
                        \lIf{$\txnVar_1 \neq \bot$}{
                            $\co' \gets \co' \cup \{ \tuple{\txnVar_2, \txnVar_1} \}$ \label{line:rc_co_update}
                        }
                    }
                }
                \If{$\earliestWriter[y][1] \neq \txnVar_2$}{ \label{line:rc_t2_latest_wt_check}
                    $\earliestWriter[y] \gets \tuple{\earliestWriter[y][1], \txnVar_2}$ \; \label{line:rc_earliest_writer_update}
                    $\readKeys \gets \readKeys \cup \{ y \}$ \; \label{line:rc_read_keys_update}
                }
            }
        }
        \lIf{$\co'$ has a cycle}{ \label{line:rc_cycle_check}
            \Report{cycle} \label{line:rc_cycle_report}
        }
    }
\end{algorithm}

In this section, we present the algorithm for checking consistency for $\RC$ (\cref{alg:rc}).

\Paragraph{Description of algorithm.}
The algorithm starts by checking the history for Read Consistency (\cref{line:rc_internal_consistency}).
Then, it initializes $\co'$ as $\so \cup \wr$ (\cref{line:rc_co_init}), which must hold for $\co'$ to be saturated.
The main part of the algorithm saturates $\co'$ according to the $\RC$ axiom (\cref{subfig:isolation-levels-rc}), by looping over all committed transactions $\txnVar_3$ (\cref{line:rc_t3_loop}).
The loop on \cref{line:rc_r_loop} iterates over each transaction $\txnVar_2$ that $\txnVar_3$ reads from, and stores in the set $\firstTxnReads$ the first read operation of $\txnVar_3$ reading from $\txnVar_2$.
The algorithm then loops over all reads $\Read{y, \valVar}$ in $\txnVar_3$ in \emph{reverse} order (\cref{line:rc_read_loop}), while maintaining the set of keys that have been read below the current read in the $\readKeys$ variable (\cref{line:rc_read_keys_update}).
This is because $\Read{y, \valVar}$ plays the role of $\readVar$ in the $\RC$ axiom (\cref{subfig:isolation-levels-rc}), hence the intersection on \cref{line:rc_x_loop} contains all $x$ such that $\txnVar_2 \relop{\wr} \Read{y, \valVar} \relop{\po} \readVar_x$, where $\txnVar_2$ is some transaction writing $x$, and $\readVar_x$ reads $x$.
To achieve the stated complexity, it is crucial to only compute this intersection once for each $\txnVar_2$, hence the check on \cref{line:rc_first_txn_check}.
By inspecting \cref{subfig:isolation-levels-rc}, it is apparent that no $\co$ edges are missed this way, since $\Read{y, \valVar}$ is the $\po$-first read of $\txnVar_2$ by $\txnVar_3$.
Any reads $\readVar_x$ $\po$-below $\Read{y, \valVar}$ reading a key $x$ from this intersection could then create a $\co$-inference: if there is $\txnVar_1 \neq \txnVar_2$ such that $\txnVar_1 \relop{\wr} \readVar_x$, we have $\txnVar_2 \relop{\co} \txnVar_1$.
However, recall that a saturated $\co'$ only needs to contain this ordering \emph{transitively} (cf. \cref{def:saturated}): $\txnVar_2 \relopT{\co'} \txnVar_1$.
Hence, it suffices to infer $\co'$ for the earliest such read (in $\po$).
In particular, consider two reads $\readVar_x$ and $\readVar_x'$ reading $x$ from $\txnVar_1$ and $\txnVar_1'$, respectively, with $\Read{y, \valVar} \relop{\po} \readVar_x \relop{\po} \readVar_x'$.
When the algorithm processes the first read of $\txnVar_1$ on \cref{line:rc_read_loop}, it infers $\txnVar_1 \relop{\co'} \txnVar_1'$, thus it only remains to infer $\txnVar_2 \relop{\co'} \txnVar_1$.
The algorithm efficiently identifies $\txnVar_1$ as follows.

The $\earliestWriter$ map (\cref{line:rc_earliest_writer_def}) maintains, for each key $x$, the two $\po$-earlieset \emph{unique} transactions from which $\txnVar_3$ reads $x$ in the future.
It is essentially a stack of two elements for each key, where a new transaction ejects the oldest writer (\cref{line:rc_t2_latest_wt_check} and \cref{line:rc_earliest_writer_update}).
When finding the transaction $\txnVar_1$ writing the value read for a key $x$, the top element of the stack is chosen (\cref{line:rc_t1_def_fst}), except if the top is equal to $\txnVar_2$, in which case the second element is used (\cref{line:rc_t1_def_snd}).
Finally, $\txnVar_2 \relop{\co'} \txnVar_1$ is added on \cref{line:rc_co_update}.

To understand the need for this two-element stack, 
suppose the algorithm instead always used the most recent transaction that $\txnVar_3$ read $x$ from.
One could have $\readVar \relop{\po} \readVar_x \relop{\po} \readVar_x'$, where $\readVar$ and $\readVar_x$ read from the \emph{same} transaction $\txnVar_2$.
In such a case, there should still be a $\co'$ ordering between $\txnVar_2$ and the writer of $\readVar_x'$, which would be missed.

The correctness of \cref{alg:rc} follows by arguing that $\co'$ is saturated and minimal, thereby applying \cref{lem:saturated}.
\begin{restatable}{lemma}{lemrcuppercorrectness}\label{lem:rc_upper_correctness}
Given a history $\historyVar$, \cref{alg:rc} reports a violation iff $\historyVar$ does not satisfy $\RC$.
\end{restatable}

\Paragraph{Running time.}
Read Consistency can be checked in linear time, so the running time is dominated by the loop on \cref{line:rc_x_loop}.
The intersection in this loop is performed by iterating over the smaller of the two sets, which leads to amortized $O(\sqrt{\numEvents})$ time.
We sketch the argument here.
Call a transaction \emph{large}, if it has more than $\sqrt{\numEvents}$ reads, and call it \emph{small} otherwise.
We count separately the total running time for small and large transactions encountered on \cref{line:rc_t3_loop}.
Note that there are $\leq \sqrt{\numEvents}$ large transactions, hence we argue that each large transaction $\txnVar_3$ takes $O(\numEvents)$ time.
This is true because $\txnVar_2$ is unique each time we enter \cref{line:rc_x_loop}, and $\sum_{\txnVar_2 \relop{\wr} \txnVar_3} |\keysWritten{\txnVar_2}| = O(\numEvents)$.
We now turn our attention to small transactions.
For each small transaction $\txnVar_3$, we have $|\readKeys| \leq |\ReadsOf{\txnVar_3}|$, hence the inner loop runs $O(|\ReadsOf{\txnVar_3}|^2)$ times.
The sum of these is maximized, when each $|\ReadsOf{\txnVar_3}| = \theta(\sqrt{\numEvents})$.
In this case there are $O(\sqrt{\numEvents})$ small transactions, yielding $O(\numEvents^{3/2})$ total time.
Finally, note that when each transaction has constant size $O(1)$, the above argument yields $O(\numEvents)$ running time.
We thus arrive at the following lemma, which concludes \cref{thm:ra_rc_upper} for $\RC$.

\begin{restatable}{lemma}{lemrcuppercomplexity}\label{lem:rc_upper_complexity}
Given a history $\historyVar$ of size $\numEvents$, \cref{alg:rc} runs in $O(\numEvents^{3/2})$ time.
\end{restatable}

\subsection{Read Atomic}\label{SUBSEC:ALGORITHMS_RA}

\begin{algorithm}
\small
    \caption{Read Atomic}
    \label{alg:ra}
    
\begin{multicols}{2}
    \Fn{\FCheckRA{$\historyVar = \tuple{\TxnSet, \so, \wr}$}}{
		\tcp{\cref{alg:internal-consistency}}
        \FCheckInternalConsistency{$\historyVar$}\; \label{line:ra_internal_consistency}
        \FCheckRepeatableReads{$\historyVar$} \; \label{line:ra_repeatable_reads}
        $\co' \gets \so \cup \wr$ \; \label{line:ra_co_def}
        \For{$\sessionVar \in \SessionsOf{\historyVar}$}{ \label{line:ra_sess_loop}
            $\lastWrite{} \gets \lambda x . \bot$ \;
            \For{$\txnVar_3 \in \TxnsOfSession{\historyVar}{\sessionVar}$ in $\so$ order}{ \label{line:ra_t3_loop}
                \For{$\txnVar_1 \relop{\wr_x} \txnVar_3$}{ \label{line:ra_t1_loop}
                    $\txnVar_2 \gets \lastWrite{}[x]$ \; \label{line:ra_t2_def}
                    \If{$\txnVar_1 \neq \txnVar_2 \neq \bot$}{
                        $\co' \gets \co' \cup \{ \tuple{\txnVar_2, \txnVar_1} \}$ \; \label{line:ra_co_update_so}
                    }
                }
                \For{$\txnVar_2 \relop{\wr} \txnVar_3$}{ \label{line:ra_t2_loop}
                    \tcp{Loop over the smaller set}
                    \For{$x \in \keysWritten{\txnVar_2} \cap \keysRead{\txnVar_3}$}{\label{line:ra_x_loop}
                        \tcp{$t_1$ is unique due to repeatable reads}
                        Let $\txnVar_1$ be such that $\txnVar_1 \relop{\wr_x} \txnVar_3$ \; \label{line:ra_t1_def}
                        \If{$t_1 \neq t_2$}{
                            $\co' \gets \co' \cup \{ \tuple{\txnVar_2, \txnVar_1} \}$ \; \label{line:ra_co_update_wr}
                        }
                    }
                }
                \For{$x \in \keysWritten{\txnVar_3}$}{\label{line:ra_last_write_loop}
                    $\lastWrite{}[x] \gets \txnVar_3$
                }
            }
        }
        
        \If{$\co'$ has a cycle}{ \label{line:ra_acyclicity_check}
            \Report{cycle} \; \label{line:ra_cycle_report}
        }
    }
    
    \Fn{\FCheckRepeatableReads{$\historyVar= \tuple{\TxnSet, \so, \wr}$}}{
        \For{$\txnVar \in \TxnSetCommitted$}{
            $\lastWriter \gets \lambda x. \bot$ \;
            \For{$\Read{x, \valVar} \in \ReadsOf{\txnVar}$ in $\po$ order}{
                \uIf{$\txnVar \neq \txnOf{\Write{x, \valVar}} \neq \lastWriter[x] \neq \bot$}{
                    \tcp{Found a cycle between the writer of $v$ and $\lastWriter[x]$}
                    \Report{non-repeatable read} \;
                }
                \Else{
                    $\lastWriter[x] \gets \txnOf{\Write{x, \valVar}}$ \;
                }
            }
        }
    }
\end{multicols}
\end{algorithm}
In this section we present the algorithm for checking consistency for $\RA$ (\cref{alg:ra}).

\Paragraph{Description.}
The algorithm for $\RA$ is similar to \cref{alg:rc} in its overall approach.
It starts by checking Read Consistency (\cref{line:ra_internal_consistency}), followed by checking the \emph{repeatable reads} property (\cref{line:ra_repeatable_reads}).
In short, repeatable reads states that committed transactions don't read the same key from different transactions, and is implied by the $\RA$ axiom (\cref{subfig:isolation-levels-ra}).
The algorithm proceeds by initializing $\co'$ (\cref{line:ra_co_def}) and then looping over all sessions $\sessionVar$ (\cref{line:ra_sess_loop}) and all committed transactions $\txnVar_3$ in $\sessionVar$ (\cref{line:ra_t3_loop}).
The algorithm maintains $\lastWrite{}$, which holds, for each key $x$, the latest transaction in $\sessionVar$ so far that writes $x$.
The $\RA$ axiom includes the condition $\txnVar_2 \relop{\so\,\cup\,\wr} \txnVar_3$, which is handled as two separate cases.
The $\so$ case is handled by the loop on \cref{line:ra_t1_loop}, whereas the $\wr$ case is handled by the loop on \cref{line:ra_t2_loop}.
For the $\so$ case, the algorithm exploits that saturation only requires \emph{transitive} $\co'$ orderings, when the $\RA$ axiom applies;
in a scenario $\txnVar_2' \relop{\so} \txnVar_2 \relop{\so} \txnVar_3$, where $\txnVar_2$ and $\txnVar_2'$ write $x$ and $\txnVar_1 \relop{\wr_x} \txnVar_3$, it is only necessary to infer $\txnVar_2 \relop{\co'} \txnVar_1$, because $\txnVar_2' \relop{\so} \txnVar_2 \relop{\co'} \txnVar_1$.
In the $\wr$ case, the algorithm iterates all possible $\txnVar_2$ (\cref{line:ra_t2_loop}) and finds exactly those keys $x$ for which the $\RA$ axiom could apply by computing an intersection (\cref{line:ra_x_loop}).
As with $\RC$, it is crucial for the complexity that this intersection is performed by iterating over the smaller set.
The inferred $\co'$ edges are added (\cref{line:ra_co_update_so} and \cref{line:ra_co_update_wr}) and $\co'$ is finally checked for cycles (\cref{line:ra_acyclicity_check}).

The complexity of \cref{alg:ra} follows a similar line of reasoning to that of $\RC$: the running time is dominated by the loop on \cref{line:ra_x_loop}, which can be shown to run in amortized $O(\sqrt{n})$ time, by again reasoning about small and large transactions separately.
Formally, the correctness and complexity of \cref{alg:ra} is captured in the following lemmas, which conclude \cref{thm:ra_rc_upper} for $\RA$.

\begin{restatable}{lemma}{lemrauppercorrectness}\label{lem:ra_upper_correctness}
Given a history $\historyVar$, \cref{alg:ra} reports a violation iff $\historyVar$ does not satisfy $\RA$.
\end{restatable}
\begin{restatable}{lemma}{lemrauppercomplexity}\label{lem:ra_upper_complexity}
Given a history $\historyVar$ of size $\numEvents$, \cref{alg:ra} runs in $O(\numEvents^{3/2})$ time.
\end{restatable}
\subsection{Causal Consistency}\label{SUBSEC:ALGORITHMS_CC}

In this section we present the algorithm for checking consistency for $\CC$ (\cref{alg:cc}).

\begin{algorithm}
\small
    \caption{Causal Consistency}
    \label{alg:cc}

\begin{multicols}{2}
    \Fn{\FCheckCC{$\historyVar = \tuple{\TxnSet, \so, \wr}$}}{
		\tcp{\cref{alg:internal-consistency}}
        \FCheckInternalConsistency{$\historyVar$}\label{line:algo_cc_read_consistency} \;
        $\HB \gets \FComputeHB{\historyVar}$ \label{line:algo_cc_call_hb} \;
        $\co' \gets \so \cup \wr$ \label{line:algo_cc_initco} \;

        \For{$\sessionVar \in \SessionsOf{\historyVar}$}{\label{line:algo_cc_main_loop}
            $\lastWrite{\sessionVar'} \gets \lambda x . \bot$ for each $\sessionVar' \in \SessionsOf{\historyVar}$ \;
            \For{$\txnVar_3 \in \TxnsOfSession{\historyVar}{\sessionVar}$ in $\so$ order}{\label{line:algo_cc_iterate_transactions}
                \For{$\txnVar_1 \relop{\wr_x} \txnVar_3$}{\label{line:algo_cc_iterate_wr}
                    \For{$\sessionVar' \in \SessionsOf{\historyVar}$}{
                        \For{$\txnVar_2 \in \Writes{\sessionVar'}[x]$ $\so^?$-after $\lastWrite{\sessionVar'}[x]$}{\label{line:algo_cc_last_write}
                            \uIf{$\txnVar_2 \relopR{\so} \HB_{\txnVar_3}[\sessionVar']$}{
                                $\lastWrite{\sessionVar'}[x] \gets \txnVar_2$ \;
                            }
                            \lElse{
                                \Break 
                            }
                        }
                        \If{$\txnVar_1 \neq \lastWrite{\sessionVar'}[x] \neq \bot$}{
                            $\co' \gets \co' \cup \tuple{\lastWrite{\sessionVar'}[x], \txnVar_1}$\label{line:algo_cc_saturate}
                        }
                    }
                }
            }
        }
        
        \lIf{$\co'$ has a cycle}{
            \Report{cycle}
        }
    }
    
    \Fn{\FComputeHB{$\historyVar= \tuple{\TxnSet, \so, \wr}$}}{
        \lIf{$\so \cup \wr$ has a cycle}{\label{line:algo_cc_hb_acyclic}
            \Report{cycle}
        }
        
        \For{$\sessionVar \in \SessionsOf{\historyVar}$}{
            $\HB^{\sessionVar}\gets [\bot,\dots,\bot]$ \;
        }
        Let $\sigma$ be a topological sort of $\so \cup \wr$ \;
        \For{$\txnVar \in \sigma$}{
            $\sessionVar \gets \SessionOf{\txnVar}$ \;
            $\HB_\txnVar \gets \HB^{\sessionVar} \sqcup \bigsqcup_{\txnVar' \relop{\wr} \txnVar} \HB_{\txnVar'}$ \label{line:algo_cc_hb_join} \;
            $\HB^{\sessionVar} \gets \HB_{\txnVar}[\sessionVar \mapsto \txnVar]$ \;
        }
        \Return{$\HB$}
    }
\end{multicols}
\end{algorithm}
\Paragraph{Description.}
The algorithm starts by checking Read Consistency (\cref{line:algo_cc_read_consistency}) and computing the happens before relation (\cref{line:algo_cc_call_hb}) by calling $\FComputeHB{\historyVar}$.
In turn, this function verifies that $\so\cup\wr$ is acyclic (\cref{line:algo_cc_hb_acyclic}) and computes the happens before relation as a set of Vector Clocks $\HB_{\txnVar}$, one for each transaction $\txnVar$.
Vector Clocks are indexed by sessions, so that for each $\sessionVar\in \SessionsOf{\historyVar}$, $\HB_{\txnVar}[s]$ holds the $\so$-latest transaction $\txnVar'$ of $\sessionVar$ such that $\txnVar'\relopT{\so\,\cup\,\wr}\txnVar$.
The join operation between two Vector Clocks $A$ and $B$ (used on \cref{line:algo_cc_hb_join}) is defined as a point-wise maximum wrt $\so$, i.e.,
\[
A\sqcup B = \lambda \sessionVar. \left( A[\sessionVar]\relop{\so}B[\sessionVar]\quad?\quad B[\sessionVar]\quad:\quad A[\sessionVar] \right)\ .
\]
The algorithm then initializes $\co'$ to $\so\cup\wr$ (\cref{line:algo_cc_initco}) and enters its main computation in the loop of \cref{line:algo_cc_main_loop}, so as to saturate $\co'$ based on selective applications of the $\CC$ axiom in \cref{subfig:isolation-levels-cc}.
This is achieved by iterating over all transactions $\txnVar_3$ of each session $\sessionVar$, in $\so$-order.
To make the computation efficient, the algorithm relies on two simple data structures.
The last-writer data structure $\lastWrite{\sessionVar'}[x]$ points to the $\so$-latest transaction $\txnVar'_2$ writing $x$ of session $\sessionVar'$ such that $\txnVar'_2\relopT{\so\,\cup\,\wr} \txnVar_3$.
In accordance with the $\CC$ axiom, for $\txnVar_2 = \lastWrite{s'}[x]$ and $\txnVar_1$, the transaction for which $\txnVar_1 \relop{\wr_x} \txnVar_3$, if $\txnVar_1\neq \txnVar_2$, we have $\txnVar_2 \relop{\co'} \txnVar_1$ (\cref{line:algo_cc_saturate}).
Importantly, the algorithm avoids inserting orderings $\txnVar_2' \relop{\co'} \txnVar_1$ from transactions $\txnVar_2'$ that are $\so$-predecessors of $\txnVar_2$, since these will be ordered before $\txnVar_1$ transitively via $\txnVar_2$.
Finally, the last-writer data structure  $\lastWrite{\sessionVar'}[x]$ is updated by traversing $\Writes{\sessionVar'}[x]$, which is an array storing the transactions of $\sessionVar'$ that write on $x$, in $\so$ order.
The key insight is that last writers grow monotonically with $\so$:~for $\txnVar_3$ and $\txnVar'_3$ of $\sessionVar$ with $\txnVar_3\relop{\so}\txnVar'_3$, we have
\[
\{\txnVar_2\in \TxnSet \mid \txnVar_2\relopT{\so\,\cup\,\wr}\txnVar_3\}\subseteq \{\txnVar_2\in \TxnSet \mid \txnVar_2\relopT{\so\,\cup\,\wr}\txnVar'_3\}\ .
\]
This implies that, after processing $\txnVar_3$ and proceeding to $\txnVar_3'$, $\lastWrite{\sessionVar'}[x]$ does not have to scan the array $\Writes{\sessionVar'}[x]$ from the beginning, but rather proceed from where it left of on $\txnVar_3$ (\cref{line:algo_cc_last_write}).

\Paragraph{Running time.}
We now sketch the running time of \cref{alg:cc}.
$\FComputeHB{\historyVar}$ clearly takes $O(\numEvents\cdot \numSessions)$, by spending $O(\numSessions)$ time per event for each join operation.
For every $\txnVar_1 \relop{\wr_x} \txnVar_3$ in \cref{{line:algo_cc_iterate_wr}}, the algorithm spends $O(\numSessions)$ time, if we exclude the inner loop of \cref{line:algo_cc_last_write}.
Finally, the total time spent per session in the inner loop of \cref{line:algo_cc_last_write} is bounded by $O(\numEvents)$, since, as argued above, each array $\Writes{\sessionVar'}[x]$ is scanned once for each transaction $s$.
We thus arrive at a total running time of $O(\numEvents \cdot \numSessions)$.

Formally, the correctness and complexity of \cref{alg:cc} is captured in the following lemmas, which conclude \cref{thm:cc_upper}.

\begin{restatable}{lemma}{lemcccorrectness}\label{lem:cc_correctness}
Given a history $\historyVar$, \cref{alg:cc} reports a violation iff $\historyVar$ does not satisfy $\CC$.
\end{restatable}

\begin{restatable}{lemma}{lemcccomplexity}\label{lem:cc_complexity}
Given a history $\historyVar$ of size $\numEvents$ and $\numSessions$ sessions,
\cref{alg:cc} runs in $O(\numEvents\cdot \numSessions)$ time.
\end{restatable}

\subsection{Witnesses of Reported Violations}\label{SUBSEC:ALGORITHMS_REPORTING}

Besides merely reporting whether a history $\historyVar$ satisfies a given isolation level, it is informative to extract witnesses of isolation anomalies.
It is also desirable, that we extract several independent anomalies that are possibly present in $\historyVar$.
Although our algorithms so far make coarse-grained reports (e.g., in terms of the existence of a $\co'$ cycle) for reasons of brevity, here we present some fine-grained witness-reporting strategies that are easily obtainable with our algorithms.

\Paragraph{Read Consistency.}
Each algorithm starts by checking Read Consistency (\cref{fig:internal-consistency}), required by all isolation levels.
Each read is checked independently, thus the algorithms can report \emph{all} reads failing one of the five basic axioms (e.g., future read or thin-air read).
Even in the presence of violations, consistency checking can still proceed, by discarding the reads that suffer an anomaly at this level.

\Paragraph{Causality cycles.}
The next weakest anomaly is the presence of causality ($\so\cup \wr$) cycles, which violates all isolation levels.
Though the number of cycles can be exponential, one can obtain meaningful witnesses by reporting one cycle per strongly connected component (SCC) of $\so\cup \wr$.
At this point, consistency checks for $\RC$ and $\RA$ (the axioms that do not involve $(\so\cup \wr)^+$) may continue, 
while consistency checks for $\CC$ is likely to produce too many violation reports.

\Paragraph{Commit-order violations.}
Next, we proceed to isolation-level-specific anomalies.
The repeatable read property of $\RA$ is checked (and reported) independently for each transaction.
All other anomalies (for all isolation levels) involve the presence of a $\co'$ cycle.
Again, we find it meaningful to report one cycle per SCC, but it is also insightful to consider the edges constituting each cycle.
One approach is to prioritize cycles that contain the fewest non-$(\so\cup\wr)$ edges, which is likely to report weaker (and thus more serious) anomalies.

\Paragraph{Reporting all violations.}
Finally, we remark that more exhaustive witness reporting is possible.
Although it may come at a higher complexity cost, it only has to execute after \emph{the first} violation is reported, which can be done optimally using the algorithms of this section.
Since the vast majority of tested histories do not have violations, this approach still benefits from our faster algorithms.

\section{Complexity Lower Bounds}\label{SEC:LOWER_BOUNDS}

In this section we turn our attention to the lower bounds of \cref{thm:range_lower}, \cref{thm:ralower}, and \cref{thm:rclower}, which broadly state that testing weak database isolation on histories of size $\numEvents$ essentially requires $\numEvents^{3/2}$ time, in the sense that polynomial improvements over this scaling are unlikely.

\Paragraph{Triangle freeness and boolean matrix multiplication.}
Triangle freeness is a simple graph-theoretic problem:
given an undirected graph $G=\tuple{V, E}$, does it contain a triangle, i.e., three nodes $a,b,c$ with $\tuple{a,b}, \tuple{b,c}, \tuple{a,c}\in E$?
Triangle freeness has been studied extensively.
It is solvable in $O(n^3)$ time, on a graph of $n$ nodes,
and, although faster algorithms exist, it is also BMM-hard~\cite{Williams2018}.
This means that any combinatorial algorithm computing triangle freeness in $O(n^{3-\epsilon})$ time would imply the existence of a combinatorial algorithm for multiplying two $n\times n$ matrices in $O(n^{3-\epsilon'})$ time, for fixed $\epsilon, \epsilon'>0$.
The latter is considered unlikely (or at least notoriously difficult).
It also implies that triangle freeness cannot be solved in $O(n^{\omega-\epsilon})$ time, where $\omega$ is the matrix multiplication exponent.

Our lower bounds are based on fine-grained reductions from triangle freeness.

\subsection{A General Lower Bound for Weak Isolation Testing}\label{SUBSEC:LOWER_BOUND_RANGE}

We begin with \cref{thm:range_lower}.
Given an undirected graph $G = \tuple{V, E}$, we construct a history $H = \tuple{T, \so, \wr}$ such that, for any isolation level $\IsolationLevel$ with $\CC\StrongerThan\IsolationLevel\StrongerThan\RC$, we have that $\historyVar$ satisfies $\IsolationLevel$ iff $G$ is triangle-free.
We achieve this by means of a \emph{range reduction}, which has the property that
(i) if $G$ is triangle-free, then $\historyVar$ satisfies $\CC$ (and thus also $\IsolationLevel$), and
(ii) if $\historyVar$ satisfies $\RC$ (and thus also $\IsolationLevel$), then $G$ is triangle-free.

\begin{figure}
    \centering
    \begin{subfigure}[b]{0.30\textwidth}
        \centering
\tikzset{every node/.style={draw, circle, thick},}
        \begin{tikzpicture}[font=\small]
            \graph[math nodes, clockwise, n=3, radius=3em]{
                1, 2, 3;
                1 -- 2 -- 3 -- 1;
            };
			\node[draw=none] at (0,-1.5) {};
        \end{tikzpicture}
        \caption{An undirected graph $G$.}
		\label{subfig:range_reduction_example_graph}
    \end{subfigure}
    \begin{subfigure}[b]{0.6\textwidth}
        \centering
	\scalebox{0.95}{%
        \begin{tikzpicture}[font=\small]
            \path (0,0) graph[nodes={transaction, text width=1.1cm}, clockwise, n=3, radius=7em]{
                1in/{
                    $\Read{x_2^1, 2}$\\
                    $\Read{x_3^1, 3}$\\
                    $\Read{x_2, 2}$\\
                    $\Read{x_3, 3}$
                }[label=180:$\txnVar_1^\R$],
                2in/{
                    $\Read{x_1^2, 1}$\\
                    $\Read{x_3^2, 3}$\\
                    $\Read{x_1, 1}$\\
                    $\Read{x_3, 3}$
                }[label=180:$\txnVar_2^\R$],
                3in/{
                    $\Read{x_1^3, 1}$\\
                    $\Read{x_2^3, 2}$\\
                    $\Read{x_1, 1}$\\
                    $\Read{x_2, 2}$
                }[label=180:$\txnVar_3^\R$];
            };

			\begin{scope}[shift={(1.5,0.18)}]
            \path (0,0) graph[nodes={transaction, text width=1.1cm}, clockwise, n=3, radius=7em]{
                1out/{
                    $\Write{x_1^2, 1}$\\
                    $\Write{x_1^3, 1}$\\
                    $\Write{x_2, 1}$\\
                    $\Write{x_3, 1}$\\
                    $\Write{x_1, 1}$
                }[label=0:$\txnVar_1^\W$],
                2out/{
                    $\Write{x_2^1, 2}$\\
                    $\Write{x_2^3, 2}$\\
                    $\Write{x_1, 2}$\\
                    $\Write{x_3, 2}$\\
                    $\Write{x_2, 2}$
                }[label=0:$\txnVar_2^\W$],
                3out/{
                    $\Write{x_3^1, 3}$\\
                    $\Write{x_3^2, 3}$\\
                    $\Write{x_1, 3}$\\
                    $\Write{x_2, 3}$\\
                    $\Write{x_3, 3}$
                }[label=0:$\txnVar_3^\W$];
            };
			\end{scope}

				\draw (1out) edge[draw=WR, ->, bend left=10] node[right, pos=0.3]{$\wr_{x_1}, \wr_{x_1^2}$} (2in);
				\draw (2out) edge[draw=WR, ->, out=140, in=-50] node[left, pos=0.8]{$\wr_{x_2}, \wr_{x_2^1}$} (1in);

				\draw (1out) edge[draw=WR, ->, out=-90, in=65] node[left, xshift=-5pt, pos=0.8]{$\wr_{x_1}, \wr_{x_1^3}$} (3in);
				\draw (3out) edge[draw=WR, ->, out=100, in=-120] node[left,  pos=0.8]{$\wr_{x_3}, \wr_{x_3^1}$} (1in);

				\draw (2out) edge[draw=WR, ->, out=-120, in=-60, looseness=0.3] node[below, pos=0.5] {$\wr_{x_2}, \wr_{x_2^3}$} (3in);
				\draw (3out) edge[draw=WR, ->, out=-30, in=200] node[below, pos=0.5] {$\wr_{x_3}, \wr_{x_3^2}$} (2in);

				\draw (3out) edge[draw=CO, ->, out=-70, in=-110, looseness=0.4] node[below, pos=0.6] {$\co$} (2out);
				\draw (2out) edge[draw=CO, ->, out=110, in=70, looseness=0.4] node[above, pos=0.6] {$\co$} (3out);
        \end{tikzpicture}
}%
        \caption{The corresponding history $\historyVar$.}
\label{subfig:range_reduction_example_history}
    \end{subfigure}
    \caption{
The history $\historyVar$ given an undirected graph $G$.
Using the semantics of $\RC$ (\cref{subfig:isolation-levels-cc}), we derive $t_3^{\W}\relop{\co}t_2^{\W}$ and  $t_2^{\W}\relop{\co}t_3^{\W}$, implying that $\historyVar$ does not satisfy $\RC$, indicating that $G$ has a triangle.
}
    \label{fig:range_reduction_example}
\end{figure}
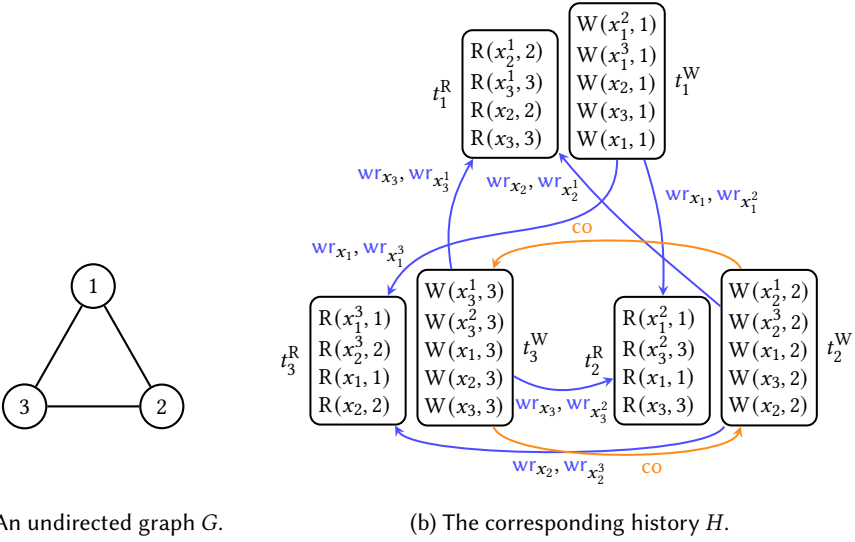

\Paragraph{Construction.}
For each node $a \in V$, $\historyVar$ has two (committed) transactions $\txnVar_a^{\R}$ and $\txnVar_a^{\W}$. 
We call the former the \emph{read transaction} and the latter the \emph{write transaction} of $a$.
\begin{compactitem}
\item The read transaction $\txnVar_a^{\R}$ begins with a sequence of reads $\Read{x_b^a,b}$, one for each edge $\tuple{b,a}\in E$.
Next (in $\po$),  $\txnVar_a^{\R}$ executes a sequence of reads $\Read{x_b, b}$, one for each edge $\tuple{b,a}\in E$.
\item The write transaction $\txnVar_a^{\W}$ contains a sequence of writes $\Write{x_b,a}$ and $\Write{x_a^b, a}$, one for each edge $\tuple{a,b}\in E$, as well as a write $\Write{x_a,a}$.
The $\po$ in $\txnVar_a^{\W}$ is irrelevant.
\end{compactitem}
Note that, for a given key, every read observes a unique value.
In particular, the $\wr$ relation is fully characterized by the following orderings.
For every edge $\tuple{a,b}\in E$, we have
(i)~$\Write{x_a^b,a} \relop{\wr}\Read{x_a^b,a}$ and
(ii)~$\Write{x_a,a} \relop{\wr}\Read{x_a,a}$,
where each write appears in $\txnVar_{a}^{\W}$ and each read appears in $\txnVar_b^{\R}$.
Finally, each transaction appears in its own session (i.e., $\so=\emptyset$).
See \cref{fig:range_reduction_example} for an illustration.

\Paragraph{Correctness.}
We now sketch the correctness of the construction.
Consider an edge $\tuple{a,b}\in E$.
The ordering $\Write{x_a^b,a} \relop{\wr}\Read{x_a^b,a}$ ensures that $\txnVar_{a}^{\W}\relop{\wr}\Read{x_a^b,a}\relop{\po}\Read{x_c,c}$, for any $\Read{x_c,c}$ of $\txnVar_b^{\R}$ and edge $\tuple{b,c}\in E$.
At this point, a triangle will be formed iff $\tuple{a,c}\in E$.
If so, then $\txnVar_a^{\W}$ also writes to $x_c$, forcing the commit order $\txnVar_a^{\W}\relop{\co}\txnVar_c^{\W}$ according to the semantics of $\RC$ (\cref{subfig:isolation-levels-rc}).
Repeating the argument symmetrically implies $\txnVar_c^{\W}\relop{\co}\txnVar_a^{\W}$, making $\historyVar$ inconsistent.
On the other hand, if there is no triangle, no $\co$ orderings are forced between write transactions, meaning that they can be committed in any order (followed by the read transactions).

\begin{example}
Let us illustrate the above argument on the example in \cref{fig:range_reduction_example}.
The edge $\tuple{3,1}$ implies $\txnVar_{3}^{\W}\relop{\wr} \Read{x_3^2,3} \relop{\po}\Read{x_2,2}$, where the reads belongs to $\txnVar_1^{\R}$ and $\Read{x_2,2}$ reads from $\Write{x_2,2}$ of $\txnVar_3^{\W}$.
Further, the edge $\tuple{3,2}$ implies that $\txnVar_3^{\W}$ also writes on $x_2$, namely via $\Write{x_2,3}$, implying a commit order $\txnVar_3^{\W}\relop{\co}\txnVar_2^{\W}$.
Exchanging nodes $3$ and $2$ and repeating this argument yields $\txnVar_2^{\W}\relop{\co}\txnVar_3^{\W}$, producing a cycle that witnesses the inconsistency of $\historyVar$ under $\RC$.
\end{example}

Formally, we have the following lemma.

\begin{restatable}{lemma}{lemrangelowercorrectness}\label{lem:range_lower_correctness}
The following assertions hold.
\begin{compactenum}
\item\label{item:range_lower1} If $G$ is triangle-free, then $\historyVar$ satisfies the $\CC$ isolation level.
\item\label{item:range_lower2} If $\historyVar$ satisfies the $\RC$ isolation level, then $G$ is triangle free.
\end{compactenum}
\end{restatable}

Finally, observe that if $G$ has $n$ nodes and $m$ edges, $\historyVar$ has size $O(m)$, which is bounded by $O(n^2)$.
Thus, if there is a combinatorial algorithm for the consistency  of $\historyVar$ in time $O(m^{3/2-\epsilon})=O(n^{3-\epsilon'})$, for some fixed $\epsilon, \epsilon'>0$,
then triangle freeness in $G$ would be determined in time $O(n^{3-\epsilon'})$, contradicting the combinatorial BMM hypothesis.
Similarly, if there is an algorithm for the consistency of $\historyVar$ in time $O(m^{\omega/2-\epsilon})=O(n^{\omega-\epsilon})$, then triangle freeness would be determined in $O(n^{\omega-\epsilon'})$.
This concludes the proof of \cref{thm:range_lower}.

\subsection{Lower Bounds with One and Two Sessions}\label{SUBSEC:LOWER_BOUND_RA}

Our algorithms for $\RA$ and $\RC$ have time complexity of $O(\numEvents^{3/2})$, even if the number of sessions is small.
It is thus natural to ask whether faster testing algorithms exist for histories with a small number of sessions.
This section proves \cref{thm:ralower} and \cref{thm:rclower}, which state that it is unlikely to break below $\numEvents^{3/2}$ for $\RA$ even with only two sessions and for $\RC$ with just one session.
The proofs are by a modification of the reduction of \cref{SUBSEC:LOWER_BOUND_RANGE}.

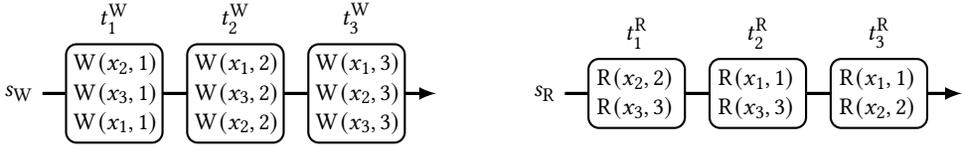
\begin{figure}
    \centering
    \newcommand{\sgap}{2}
    \newcommand{\tgap}{0.3}
    \newcommand{\slen}{5.55}
    \begin{tikzpicture}[font=\small]
        \node[] (s1) at (0, 0*\sgap) {\small$\sessionVar_\R$};
        \node[] (s2) at (-7, 0*\sgap) {\small$\sessionVar_\W$};
        \draw[session] (s1) -- ++(\slen, 0);
        \draw[session] (s2) -- ++(\slen, 0);

        \begin{scope}[]
        \node[transaction, label={\small$\txnVar_1^\R$}, right = \tgap of s1] (t1in) {
            $\Read{x_2, 2}$\\
            $\Read{x_3, 3}$
        };
        \node[transaction, label={\small$\txnVar_2^\R$}, right = \tgap of t1in] (t2in) {
            $\Read{x_1, 1}$\\
            $\Read{x_3, 3}$
        };
        \node[transaction, label={\small$\txnVar_3^\R$}, right = \tgap of t2in] (t3in) {
            $\Read{x_1, 1}$\\
            $\Read{x_2, 2}$
        };
        
        \node[transaction, label={\small$\txnVar_1^\W$}, right = \tgap of s2] (t1out) {
            $\Write{x_2, 1}$\\
            $\Write{x_3, 1}$\\
            $\Write{x_1, 1}$
        };
        \node[transaction, label={\small$\txnVar_2^\W$}, right = \tgap of t1out] (t2out) {
            $\Write{x_1, 2}$\\
            $\Write{x_3, 2}$\\
            $\Write{x_2, 2}$
        };
        \node[transaction, label={\small$\txnVar_3^\W$}, right = \tgap of t2out] (t3out) {
            $\Write{x_1, 3}$\\
            $\Write{x_2, 3}$\\
            $\Write{x_3, 3}$
        };
        \end{scope}
    \end{tikzpicture}
    \caption{The $\RA$-inconsistent history $\historyVar$ for the undirected graph $G$ of \cref{subfig:range_reduction_example_graph}, consisting of two sessions.}
    \label{fig:ra-reduction-example}
\end{figure}

\Paragraph{Reduction for $\RA$.}
The transactions of $\historyVar$ are the same as in \cref{SUBSEC:LOWER_BOUND_RANGE}, except that operations on keys $x_a^b$ are removed.
Formally, for every node $a\in V$, $\historyVar$ has two transactions $\txnVar_a^\R$ and $\txnVar_a^\W$.
\begin{compactitem}
    \item For each edge $\tuple{b,a} \in E$, the read transaction $\txnVar_a^\R$ executes a read $\Read{x_b,b}$. The $\po$ order of these is irrelevant.
    \item For each edge $\tuple{a,b} \in E$, the write transaction $\txnVar_a^\W$ executes a write $\Write{x_b,a}$. Finally, it executes a write $\Write{x_a, a}$. The $\po$ order is, again, irrelevant.
\end{compactitem}
Note that $\wr$ is fully specified by the orderings $\Write{x_a,a} \relop{\wr} \Read{x_a, a}$ for each edge $\tuple{a,b}\in E$, where $\Write{x_a,a}$ is an operation of $\txnVar_a^\W$ and $\Read{x_a, a}$ is an operation of $\txnVar_b^\R$.
The session order  $\so$ consists of two sessions $\sessionVar_\W$ and $\sessionVar_\R$, executing all write and read transactions, respectively, in some arbitrary order.
It can be easily verified, that $\so \cup \wr$ is  acyclic.
See \cref{fig:ra-reduction-example} for an illustration.

\Paragraph{Correctness for $\RA$.}
The correctness of the construction can be intuitively stated as follows.
First, the existence of a triangle $\tuple{a,b,c}$ implies the commit orderings $\txnVar_a^{\W}\relop{\co}\txnVar_c^{\W}$ and $\txnVar_c^{\W}\relop{\co}\txnVar_a^{\W}$, witnessing the inconsistency of $\historyVar$.
This holds, because the existence of edges $\tuple{a,b}, \tuple{c,b}\in E$ implies that $\txnVar_a^{\W}\relop{\wr}\txnVar_b^{\R}$ and $\txnVar_c^{\W}\relop{\wr}\txnVar_b^{\R}$, while the existence of the edge $\tuple{a,c}\in E$ implies that both $\txnVar_a^{\W}$ and $\txnVar_c^{\W}$ write the key of the other, i.e., $x_c$ and $x_a$, respectively.
On the other hand, if there is no triangle, no such $\co$ orderings are imposed.
The fact that all write and read transactions can be grouped into two sessions $\sessionVar_\W$ and $\sessionVar_\R$ follows by inspecting the $\RA$ axiom (\cref{subfig:isolation-levels-ra}):~$\so$ is irrelevant, since any transaction that reads ($t_3$ in \cref{subfig:isolation-levels-ra}) is $\so$-unordered with any transaction that writes ($t_1$ in \cref{subfig:isolation-levels-ra}).
Thus, $\historyVar$ is consistent by first committing all transactions in $\sessionVar_\W$ and then all transactions in $\sessionVar_\R$.
Formally, we have the following lemma, which concludes \cref{thm:ralower}.

\begin{restatable}{lemma}{lemralowercorrectness}\label{lem:ra_lower_correctness}
$\historyVar$ satisfies the $\RA$ isolation level iff $G$ is triangle-free.
\end{restatable}

\Paragraph{Read Atomic with one session.}
A natural question is whether a $\numEvents^{3/2}$ lower bound holds for $\RA$ with a single session.
The answer is no, as consistency in this case is checkable optimally in linear time, by scanning the single session once and keeping track of the most recent transaction writing to each location.
This concludes \cref{thm:ra_one_session_linear_time}.

\Paragraph{Reduction for $\RC$.}
Finally, we turn our attention to the reduction for $\RC$.
The construction is the same as in \cref{SUBSEC:LOWER_BOUND_RANGE}, except that we place all transactions in one session (first write transactions, followed by the read transactions).
The existence of a triangle $\tuple{a,b,c}$, again, implies two conflicting commit orderings $\txnVar_a^{\W}\relop{\co}\txnVar_c^{\W}$ and $\txnVar_c^{\W}\relop{\co}\txnVar_a^{\W}$, by following exactly the same argument as in \cref{thm:range_lower}.
On the other hand, the absence of a triangle implies that $\historyVar$ is consistent, since no additional $\co$ orderings are imposed,
and the axiom of $\RC$ (\cref{subfig:isolation-levels-rc}) does not involve $\so$ (thus, trivially $\co=\so$ in this case).
Formally, we have the following lemma, which concludes \cref{thm:rclower}.

\begin{restatable}{lemma}{lemrclowercorrectness}\label{lem:rc_lower_correctness}
$\historyVar$ satisfies the $\RC$ isolation level iff $G$ is triangle-free.
\end{restatable}
\section{Implementation and Experiments}\label{SEC:EXPERIMENTS}

In this section we report on an implementation of our algorithms in \cref{SEC:ALGORITHMS} as a tool, and on an experimental evaluation of its performance against existing weak isolation testers.

\Paragraph{The $\ToolName$ weak isolation tester.}
$\ToolName$ (\underline{A} \underline{W}eak \underline{D}atabase \underline{I}solation \underline{T}ester)\footnote{Available at \url{https://github.com/lassemoldrup/AWDIT}.} is a Rust implementation of our algorithms for testing weak isolation levels from \cref{SEC:ALGORITHMS}.
For $\CC$, the implementation differs from \cref{alg:cc} by computing $\HB$ on the fly and replacing $\lastWrite{}$ with binary search, which we found performed better.
It parses database transaction histories in various formats also used by other isolation testers such as  Plume~\cite{Liu2024a}, PolySI~\cite{Huang2023b}, DBCop~\cite{Biswas2019}, and Cobra~\cite{Tan2020}.
Finally, $\ToolName$ follows witness-reporting strategies close to those described in \cref{SUBSEC:ALGORITHMS_REPORTING}.

\subsection{Experimental Setup}\label{SUBSEC:EXPERIMENTAL_SETUP}

The goal of our experiments is to shed light on the efficiency of existing database isolation testers, and how $\ToolName$ performs in comparison.
For this reason, we have followed the experimental setup of recent literature on database isolation testing~\cite{Liu2024a,Biswas2019,Huang2023b,Tan2020}, relying on databases and benchmarks utilized in those works.

\Paragraph{Databases.}
We make use of the following databases.
\begin{compactitem}
\item PostgreSQL 17.0, a popular relational database~\cite{Postrgres}.
\item CockroachDB 24.2.4, a relational database that achieves high availability~\cite{CockroachDB}.
We ran this database as a cluster of three local replicas.
\item RocksDB 5.15.10, a fast key-value database~\cite{RocksDB}.
\end{compactitem}

\Paragraph{Benchmarks.}
In order to simulate realistic client interaction with the aforementioned databases, we use the following benchmarks.
\begin{compactitem}
\item TPC-C, an online transaction processing (OLTP) benchmark~\cite{TPCC}.
\item C-Twitter, a benchmark from the Cobra framework~\cite{Tan2020} simulating the handling of real-time big data at Twitter~\cite{Twitter}.
\item RUBiS, an auction site benchmark modeled after eBay~\cite{Amza2002}.
\end{compactitem}

\Paragraph{History generation.}
A concrete history is generated by specifying
a database and a benchmark, as well as some benchmark-specific parameters such as the number of sessions and the number of transactions.
We rely on the framework of Cobra~\cite{Tan2020} for this task, which configures each databases to provide strong transaction isolation.
Each history is then given as input to an isolation tester in its respective format, together with an isolation level.

\Paragraph{Weak isolation testers.}
We use the following database isolation testers, covering recent literature.
\begin{compactitem}
\item $\ToolName$, developed in this work.
\item Plume, the most recent and optimized weak isolation tester~\cite{Liu2024a} that supports $\RC$, $\RA$, and $\CC$.
Implemented in Java, Plume utilizes Vector Clocks~\cite{Friedemann1989} (like $\ToolName$) and Tree Clocks~\cite{Mathur2022} to efficiently compute a valid commit order, and was shown to significantly outperform all existing testers.
\item DBCop, a polynomial-time tester~\cite{Biswas2019} that supports $\CC$, implemented in Rust.
\item CausalC+, a Datalog-based tester for $\CC$~\cite{Liu2024a,Zennou2022}.
\item TCC-Mono, a MonoSAT-based tester for $\CC$ based on monotonic SMTs~\cite{Liu2024a,Bayless2015}.
\item PolySI, a MonoSAT-based tester for Snapshot Isolation ($\SIiso$)~\cite{Huang2023b}.
\end{compactitem}

For CausalC+ and TCC-Mono, we used implementations from the experimental setup of \cite{Liu2024a}.
Note that, since $\SIiso \StrongerThan \RC, \RA, \CC$, PolySI can be used to make complete (but possibly unsound) reports of weak-isolation anomalies.
Finally, we have excluded Elle~\cite{Kingsbury2020} from our experiments because it is generally unsound (its sound ``list-append'' mode is not applicable here).
Our experiments are run on an Ubuntu 22.04 machine with a second-generation \SI{2.3}{GHz} AMD Epyc CPU and \SI{64}{GB} of memory.
\subsection{Small-Scale Experiments}\label{SUBSEC:EXPERIMENTAL_PRELIM}

Since different isolation testers have different complexity guarantees (from polynomial to exponential), we perform a preliminary set of small-scale experiments to obtain an indication for the scalability of each tester.

\Paragraph{Setup.}
We generate histories using all three benchmarks and by scaling the number of transactions within a small range $[2^{10}, 2^{15}]$, while keeping the number of sessions fixed at 50.
We execute Plume, DBCop, and $\ToolName$ at the $\CC$ isolation level (recall that Causal+ and TCC-Mono run at $\CC$ by default, while PolySI runs at $\SIiso$).
We set  a timeout of 10 minutes for processing each history.

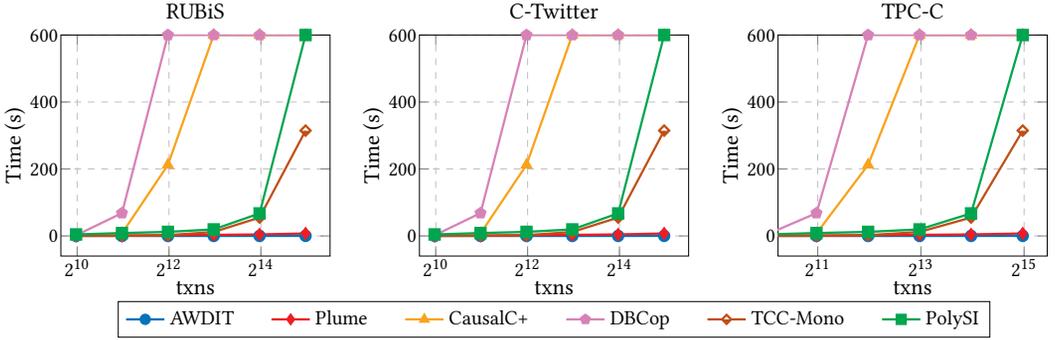
\begin{figure}
\def\scaleboxvalue{0.85}
\def\plotheight{5cm}
\def\markersize{2.25}

\pgfplotsset{AllToolsStyle/.style={
inner sep=1pt,
	title={\large Time (s)},
	mark size=\markersize,
	height=\plotheight,
	legend style={legend pos=north west, draw=black, legend columns=6, legend cell align={left}, font=\small},
	legend style={/tikz/every even column/.append style={column sep=0.5cm}},
	xlabel={txns},
    ylabel={Time (s)},
	xmin={800},
	ymax={600},
	log basis x={2},
	xtick distance = 4,
    xlabel style={font=\normalfont},
    ylabel style={font=\normalfont},
    ylabel near ticks,
    xlabel near ticks,
    xticklabel style={font=\small},
    yticklabel style={font=\small},
    grid style=dashed,
	grid=both,
	legend to name=named}
}
\def\lineWidth{1}

\pgfplotsset{
    discard if not/.style 2 args={
        x filter/.code={
            \edef\tempa{\thisrow{#1}}
            \edef\tempb{#2}
            \ifx\tempa\tempb
            \else
                \def\pgfmathresult{inf}
            \fi
        }
    }
}
\scalebox{\scaleboxvalue}{%
\begin{tikzpicture}[tight background, inner sep=0pt]
\begin{groupplot}[
group style={group size= 3 by 1, horizontal sep=1.4cm} ,
]

\def\benchmark{rubis}

\nextgroupplot[AllToolsStyle, xmode=log, title=RUBiS]

\addplot[color=RoyalBlue, line width=\lineWidth, mark=*, discard if not={script}{\benchmark} ] table [x=txns, y=ours (s), col sep=comma,] {data/2025-03-15-fig7.csv};

\addplot[color=Red, line width=\lineWidth, mark=diamond*, discard if not={script}{\benchmark} ] table [x=txns, y=plume (s), col sep=comma,] {data/2025-03-15-fig7.csv};

\addplot[color=YellowOrange, line width=\lineWidth, mark=triangle*, discard if not={script}{\benchmark} ] table [x=txns, y=dbcop (s), col sep=comma,] {data/2025-03-15-fig7.csv};

\addplot[color=Thistle, line width=\lineWidth, mark=pentagon*, discard if not={script}{\benchmark} ] table [x=txns, y=causalc+ (s), col sep=comma,] {data/2025-03-15-fig7.csv};

\addplot[color=Bittersweet, line width=\lineWidth, mark=halfsquare*, discard if not={script}{\benchmark} ] table [x=txns, y=mono (s), col sep=comma,] {data/2025-03-15-fig7.csv};

\addplot[color=Green, line width=\lineWidth, mark=square*, discard if not={script}{\benchmark} ] table [x=txns, y=polysi (s), col sep=comma,] {data/2025-03-15-fig7.csv};

\coordinate (top) at (rel axis cs:0,1);

\def\benchmark{twitter}
\nextgroupplot[AllToolsStyle, xmode=log, title=C-Twitter]

\addplot[color=RoyalBlue, line width=\lineWidth, mark=*, discard if not={script}{\benchmark} ] table [x=txns, y=ours (s), col sep=comma,] {data/2025-03-15-fig7.csv};

\addplot[color=Red, line width=\lineWidth, mark=diamond*, discard if not={script}{\benchmark} ] table [x=txns, y=plume (s), col sep=comma,] {data/2025-03-15-fig7.csv};

\addplot[color=YellowOrange, line width=\lineWidth, mark=triangle*, discard if not={script}{\benchmark} ] table [x=txns, y=dbcop (s), col sep=comma,] {data/2025-03-15-fig7.csv};

\addplot[color=Thistle, line width=\lineWidth, mark=pentagon*, discard if not={script}{\benchmark} ] table [x=txns, y=causalc+ (s), col sep=comma,] {data/2025-03-15-fig7.csv};

\addplot[color=Bittersweet, line width=\lineWidth, mark=halfsquare*, discard if not={script}{\benchmark} ] table [x=txns, y=mono (s), col sep=comma,] {data/2025-03-15-fig7.csv};

\addplot[color=Green, line width=\lineWidth, mark=square*, discard if not={script}{\benchmark} ] table [x=txns, y=polysi (s), col sep=comma,] {data/2025-03-15-fig7.csv};

\def\benchmark{tpcc}
\nextgroupplot[AllToolsStyle, xmode=log, xmin={1200}, title=TPC-C]

\addplot[color=RoyalBlue, line width=\lineWidth, mark=*, discard if not={script}{\benchmark} ] table [x=txns, y=ours (s), col sep=comma,] {data/2025-03-15-fig7.csv};

\addplot[color=Red, line width=\lineWidth, mark=diamond*, discard if not={script}{\benchmark} ] table [x=txns, y=plume (s), col sep=comma,] {data/2025-03-15-fig7.csv};

\addplot[color=YellowOrange, line width=\lineWidth, mark=triangle*, discard if not={script}{\benchmark} ] table [x=txns, y=dbcop (s), col sep=comma,] {data/2025-03-15-fig7.csv};

\addplot[color=Thistle, line width=\lineWidth, mark=pentagon*, discard if not={script}{\benchmark} ] table [x=txns, y=causalc+ (s), col sep=comma,] {data/2025-03-15-fig7.csv};

\addplot[color=Bittersweet, line width=\lineWidth, mark=halfsquare*, discard if not={script}{\benchmark} ] table [x=txns, y=mono (s), col sep=comma,] {data/2025-03-15-fig7.csv};

\addplot[color=Green, line width=\lineWidth, mark=square*, discard if not={script}{\benchmark} ] table [x=txns, y=polysi (s), col sep=comma,] {data/2025-03-15-fig7.csv};

\coordinate (bot) at (rel axis cs:1,0);

\legend{$\ToolName$, Plume, CausalC+, DBCop, TCC-Mono, PolySI }

\end{groupplot}
\path (top)--(bot) coordinate[midway] (group center);
\node[inner sep=0pt, below=0.15em] at (group center |- current bounding box.south) {\pgfplotslegendfromname{named}};
\end{tikzpicture}
}
\caption{\label{fig:all_tools_small}
Running times of all isolation testers for checking Causal Consistency on histories collected from CockroachDB, on three benchmarks (RUBiS, C-Twitter, and TPC-C), using 50 sessions.
The timeout is set to 10m.
}
\end{figure}
\Paragraph{Results.}
\cref{fig:all_tools_small} shows the results for the three benchmarks running on CockroachDB, using 50 sessions and varying the number of transactions.
DBCop, PolySI, CausalC+ and PolySI scale poorly, while $\ToolName$ and Plume run almost instantaneously.
This is in alignment with the experimental observations in~\cite{Liu2024a}, which identified scalability as one of the main challenges in weak isolation testing.
Similar observations hold with other databases and input parameters.
Given this clear difference, we only compare $\ToolName$ and Plume on large-scale experiments.

\subsection{Large-Scale Experiments}\label{subsec:experimental_all}

We now focus on the scalability of $\ToolName$ and Plume in more detail, by performing large-scale experiments across various parameters.

\Paragraph{Setup.}
We gather histories by running all three benchmarks on all three databases.
Each history consists of either $50$ or $100$ sessions, while we scale the number of transactions in the range $[2^{10}, 2^{20}]$.
This results in 198 histories in total.
We set a time out of two hours for processing each history

\begin{figure}
\def\scaleboxvalue{0.85}
\def\plotheight{5cm}
\def\markersize{2.25}
\def\xmax{7200}
\def\ymin{\xmin}
\def\ymax{\xmax}
\def\xmin{0.5e-2}
\newcommand{\DrawSpeedupLines}{
\draw[color=gray, ultra thick] (axis cs:\xmin,\ymin) -- (axis cs:\xmax,\ymax);

\draw[color=gray, very thick] (axis cs:\xmin,10*\ymin) -- (axis cs:\xmax/10,\ymax);
\draw[color=gray, very thick] (axis cs:10*\xmin,\ymin) -- (axis cs:\xmax,\ymax/10);

\draw[color=gray, thick] (axis cs:\xmin,100*\ymin) -- (axis cs:\xmax/100,\ymax);
\draw[color=gray, thick] (axis cs:100*\xmin,\ymin) -- (axis cs:\xmax,\ymax/100);

\draw[color=gray] (axis cs:\xmin,1000*\ymin) -- (axis cs:\xmax/1000,\ymax);
\draw[color=gray] (axis cs:1000*\xmin,\ymin) -- (axis cs:\xmax,\ymax/1000);
}

\pgfplotsset{ScatterStyle/.style={
inner sep=1pt,
	mark size=\markersize,
	height=\plotheight,
	legend style={legend pos=north west, draw=black, legend columns=1, legend cell align={left}, font=\small},
	xlabel={$\ToolName$ (s)},
    ylabel={Plume (s)},
    xlabel style={font=\normalfont},
    ylabel style={font=\normalfont},
    ylabel near ticks,
    xlabel near ticks,
	xtick distance=10,
	ytick distance=10,
    xmin=\xmin, xmax=\xmax,
    ymin=\ymin, ymax=\ymax,
    xticklabel style={font=\small},
    yticklabel style={font=\small},
    grid style=dashed,
	}
}
\scalebox{\scaleboxvalue}{
\begin{tikzpicture}[tight background, inner sep=0pt,]
\begin{groupplot}[
group style={group size= 3 by 1, horizontal sep=1.3cm} 
]

\nextgroupplot[ScatterStyle, xmode=log,ymode=log, title=Read Committed]

\addplot[color=RoyalBlue, only marks, mark=* ] table [x=ours_rc (s), y=plume_rc (s), col sep=comma, restrict expr to domain={\thisrow{plume_rc (s)}}{1e-2:7199}] {data/2025-03-18-fig8.csv};
\addplot[color=Red, only marks, mark=diamond* ] table [x=ours_rc (s), y=plume_rc (s), col sep=comma, restrict expr to domain={\thisrow{plume_rc (s)}}{7200:7200}] {data/2025-03-18-fig8.csv};

\DrawSpeedupLines

\nextgroupplot[ScatterStyle, xmode=log,ymode=log, title=Read Atomic]

\addplot[color=RoyalBlue, only marks, mark=* ] table [x=ours_ra (s), y=plume_ra (s), col sep=comma, restrict expr to domain={\thisrow{plume_ra (s)}}{1e-2:7199}] {data/2025-03-18-fig8.csv};
\addplot[color=Red, only marks, mark=diamond* ] table [x=ours_ra (s), y=plume_ra (s), col sep=comma, restrict expr to domain={\thisrow{plume_ra (s)}}{7200:7200}] {data/2025-03-18-fig8.csv};

\DrawSpeedupLines

\nextgroupplot[ScatterStyle, xmode=log,ymode=log, title=Causal Consistency]

\addplot[color=RoyalBlue, only marks, mark=* ] table [x=ours_cc (s), y=plume_cc (s), col sep=comma, restrict expr to domain={\thisrow{plume_cc (s)}}{1e-2:7199}] {data/2025-03-18-fig8.csv};
\addplot[color=Red, only marks, mark=diamond* ] table [x=ours_cc (s), y=plume_cc (s), col sep=comma, restrict expr to domain={\thisrow{plume_cc (s)}}{7200:7200}] {data/2025-03-18-fig8.csv};

\DrawSpeedupLines

\end{groupplot}
\end{tikzpicture}
}
\caption{\label{fig:scatter}
An aggregate performance comparison of Plume vs $\ToolName$ across all histories, for each weak isolation level.
A red diamond indicates a timeout (\SI{2}{h}) for Plume.
Gray lines mark regions of the plot indicating speedup/slowdown of $10^{i}\times$, for $i\in \{0,1,2,3\}$.
}
\end{figure}
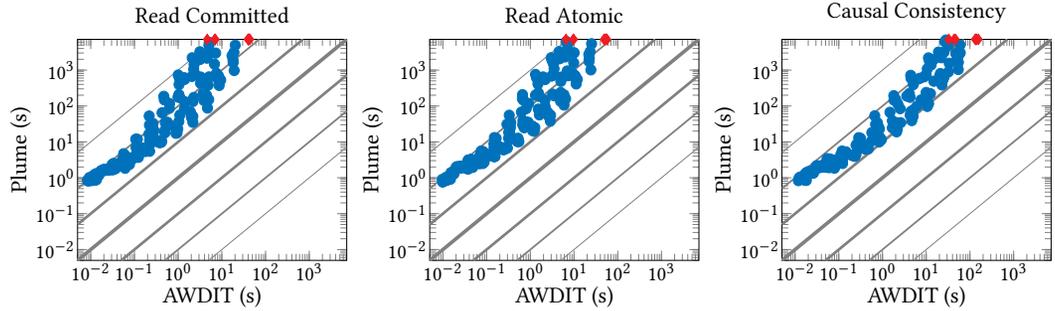

\Paragraph{Results.}
\cref{fig:scatter} shows the aggregate performance of $\ToolName$ and Plume across all histories gathered in the above setup.
We see that $\ToolName$ has a clear performance advantage, with average speedups on big histories of $245\times$, $193\times$, and $62\times$ for $\RC$, $\RA$, and $\CC$, respectively, exceeding $1000\times$ in the most extreme cases.
These averages are calculated by taking the geometric mean of the $\sim$$20\%$ largest histories (by transaction count).
The average speedups across all histories are $80\times$, $70\times$, and $36\times$ for $\RC$, $\RA$, and $\CC$, respectively.
Plume starts with a construction phase that builds a certain dependency graph for a given history, which dominates its running time for non-demanding inputs (i.e., towards the left end of the plot in each figure).
This is consistent with prior measurements~\cite{Liu2024a}, which showed that Plume's running time is often determined by the construction phase, and explains why the speedup appears to decrease initially in each figure.
However, as we move towards more demanding histories (towards the right end of the plot in each figure), Plume's solving time becomes dominant, and the speedup of $\ToolName$ increases.
Finally, Plume times out after 2 hours while analyzing a few histories, while the maximum time of $\ToolName$ is in the order of a few minutes (in particular, $\leq 2$ minutes  for $\RC$ and $\RA$, and $\leq 6$ minutes for $\CC$).

\subsection{Scalability Experiments}\label{SUBSEC:EXPERIMENTAL_SCALABILITY}

To understand the parameters that affect the running time of $\ToolName$, we turn our attention to scalability experiments.

\Paragraph{Setup.}
We consider the following three scaling settings.
\begin{compactitem}
\item Increasing the number of transactions, while keeping the number of sessions fixed at 100 and the size of each transaction bounded.
This also increases the size of each history (i.e., $\numEvents$).
\item Increasing the number of sessions (i.e., $\numSessions$), while keeping the number of transactions fixed at $10^5$ and the size of each transaction bounded.
This keeps the size of each history constant.
\item Increasing the size of each transaction, while keeping the size of each history and the number of sessions fixed at $10^6$ and 100, respectively.
\end{compactitem}
We use CockroachDB in all three settings, running the C-Twitter benchmark for the first two, which average 7.6 operations per transaction.
The last setting is not possible within the C-Twitter benchmark (it does not allow scaling the size of transactions).
Instead, we rely on a custom benchmark from the Cobra framework~\cite{Tan2020}.

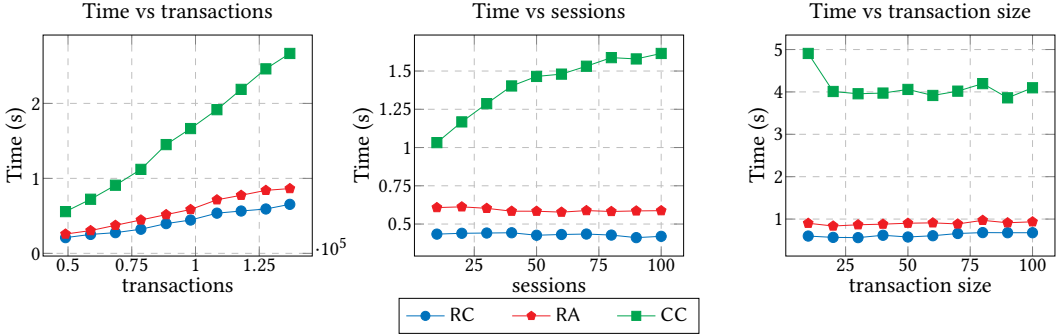
\begin{figure}
\def\scaleboxvalue{0.85}
\def\plotheight{5cm}
\def\markersize{2.25}

\pgfplotsset{ScalabilityStyle/.style={
inner sep=1pt,
	mark size=\markersize,
	height=\plotheight,
	legend style={legend pos=north west, draw=black, legend columns=6, legend cell align={left}, font=\small},
	legend style={/tikz/every even column/.append style={column sep=0.5cm}},
  ylabel={Time (s)},
	log basis x={2},
    xlabel style={font=\normalfont},
    ylabel style={font=\normalfont},
    ylabel near ticks,
    xlabel near ticks,
    xticklabel style={font=\small},
    yticklabel style={font=\small},
    grid style=dashed,
	grid=both,
	xminorgrids=false,
	yminorgrids=false,
	legend to name=named}
}
\def\lineWidth{1}

\pgfplotsset{
    discard if not/.style 2 args={
        x filter/.append code={
            \edef\tempa{\thisrow{#1}}
            \edef\tempb{#2}
            \ifx\tempa\tempb
            \else
                \def\pgfmathresult{inf}
            \fi
        }
    }
}

\def\script{twitter}
\def\database{cockroachdb}

\scalebox{\scaleboxvalue}{%
\begin{tikzpicture}[tight background, inner sep=0pt]
\begin{groupplot}[
group style={group size= 3 by 1, horizontal sep=1.6cm} ,
]

\nextgroupplot[ScalabilityStyle, xlabel={transactions}, xtick distance=25e3, ytick distance=1, every x tick scale label/.style={at={(1,0)},xshift=1pt,anchor=south west,inner sep=0pt}, title=Time vs transactions]
\addplot[color=RoyalBlue, mark=*, discard if not={script}{\script}, discard if not={database}{\database}] table [x=txns, y=ours_rc (s), col sep=comma,] {data/2025-03-18-fig9-txn.csv};
\addplot[color=Red, mark=pentagon*, discard if not={script}{\script}, discard if not={database}{\database} ] table [x=txns, y=ours_ra (s), col sep=comma,] {data/2025-03-18-fig9-txn.csv};
\addplot[color=Green, mark=square*, discard if not={script}{\script}, discard if not={database}{\database} ] table [x=txns, y=ours_cc (s), col sep=comma,] {data/2025-03-18-fig9-txn.csv};

\coordinate (top) at (rel axis cs:0,1);

\nextgroupplot[ScalabilityStyle, xlabel={sessions}, xtick distance=25, ytick distance=0.25, title=Time vs sessions]
\addplot[color=RoyalBlue, mark=*, discard if not={script}{\script}, discard if not={database}{\database}] table [x=sessions, y=ours_rc (s), col sep=comma,] {data/2025-03-18-fig9-sess.csv};
\addplot[color=Red, mark=pentagon*, discard if not={script}{\script}, discard if not={database}{\database} ] table [x=sessions, y=ours_ra (s), col sep=comma,] {data/2025-03-18-fig9-sess.csv};
\addplot[color=Green, mark=square*, discard if not={script}{\script}, discard if not={database}{\database} ] table [x=sessions, y=ours_cc (s), col sep=comma,] {data/2025-03-18-fig9-sess.csv};

\nextgroupplot[ScalabilityStyle, xlabel={transaction size}, xtick distance=25, ytick distance=1,  title=Time vs transaction size]
\addplot[color=RoyalBlue, mark=*, discard if not={script}{cheng}, discard if not={database}{\database}] table [x=ops_per_txn, y=ours_rc (s), col sep=comma,] {data/2025-03-18-fig9-ops.csv};
\addplot[color=Red, mark=pentagon*, discard if not={script}{cheng}, discard if not={database}{\database} ] table [x=ops_per_txn, y=ours_ra (s), col sep=comma,] {data/2025-03-18-fig9-ops.csv};
\addplot[color=Green, mark=square*, discard if not={script}{cheng}, discard if not={database}{\database} ] table [x=ops_per_txn, y=ours_cc (s), col sep=comma,] {data/2025-03-18-fig9-ops.csv};

\coordinate (bot) at (rel axis cs:1,0);

\legend{$\RC$, $\RA$, $\CC$}

\end{groupplot}
\path (top)--(bot) coordinate[midway] (group center);
\node[inner sep=0pt, below=0.15em] at (group center |- current bounding box.south) {\pgfplotslegendfromname{named}};
\end{tikzpicture}
}
\caption{\label{fig:scalability_awdit}
Scalability experiments on $\ToolName$ as a function of the number of transactions (left), number of sessions (middle), and number of operations per transaction (right), for each isolation level.
}
\end{figure}
\Paragraph{Results.}
\cref{fig:scalability_awdit} shows the scalability of $\ToolName$.
We see a linear effect of the number of transactions on running time (left), for each isolation level.
This aligns with our theoretical analysis, which shows a linear dependency on $\numEvents$ when the size of each transaction is bounded (for $\RC$ and $\RA$), and when the number of sessions is fixed (for $\CC$).
The slope of each curve (for each isolation level) depends on the number of sessions and the size of each transaction.
Normally, we expect the latter to be smaller, which is also the case in our experiments, resulting in a smaller slope for $\RA$ and $\RC$.

Next, we turn our attention to scaling the number of sessions $\numSessions$ (middle).
We observe an increase in the running time of $\ToolName$ for $\CC$, again in alignment with our theoretical analysis,
which predicts a cost of $\numSessions$ on average for each operation in the history.
On the other hand, increasing the number of sessions has no effect on the running time of $\ToolName$ for $\RC$ and $\RA$, which are only affected by $\numEvents$ and the size of each transaction (both of which are bounded in this case).

Finally, we turn our attention to scaling the size of each transaction (right).
Here, we observe no discernible scaling for any of the isolation levels.
This is as predicted for $\CC$, whereas our worst-case analysis predicts scaling for $\RC$ and $\RA$, as transaction size approaches $\sqrt{n}$.
This indicates that the $\RC$ and $\RA$ algorithms exhibit near-linear scaling on a variety of inputs.

\subsection{Isolation Anomalies Detected}\label{SUBSEC:EXPERIMENTS_ANOMALIES}

We have verified that $\ToolName$ and Plume agree on their reports of inconsistent histories.
Naturally, this only holds for histories that Plume does not time out.
In total, $\ToolName$ finds isolation anomalies on $8$ histories across all our experiments, summarized in \cref{tab:violations}.
Plume misses the anomaly in $\historyVar_8$ due to a timeout (after two hours) and also misses the anomalies in $\historyVar_2$ and $\historyVar_4$ when run on the $\RA$ and $\CC$ isolation levels, due to a timeout (after 10 minutes) and a crash, respectively.

\begin{table}
\setlength\tabcolsep{2.3pt}
\renewcommand{\arraystretch}{0.9}
\caption{Isolation anomalies reported by $\ToolName$ and Plume.}
\label{tab:violations}
\begin{tabularx}{\textwidth}{@{}r ccccc cc@{}}
\toprule
&
\multicolumn{4}{c}{\textbf{Parameters}} &
&
\multicolumn{2}{c}{\textbf{Reported?}} \\

\cmidrule(lr){2-5}
\cmidrule(lr){7-8}

\textbf{History} & \textbf{Size} & \textbf{Sessions} & \textbf{Database} & \textbf{Benchmark} & \textbf{Violation(s)} & \textbf{AWDIT} & \textbf{Plume} \\ \midrule
$\historyVar_1$ & 32768 & 100 & CockroachDB & TPC-C & Future Read & \cmarkgreen & \cmarkgreen \\ \midrule
$\historyVar_2$ & 50000 & 30 & CockroachDB & TPC-C & \begin{tabular}[c]{@{}c@{}}Future Read\\ Causality Cycle\end{tabular} & \cmarkgreen & \begin{tabular}[c]{@{}c@{}}\cmarkgreen / \xmarkred\\ \small(only in $\RC$)\end{tabular} \\ \midrule
$\historyVar_3$ & 2048 & 50 & PostgreSQL & TPC-C & Future Read & \cmarkgreen & \cmarkgreen \\ \midrule
$\historyVar_4$ & 16384 & 50 & PostgreSQL & TPC-C & \begin{tabular}[c]{@{}c@{}}Future Read\\ Causality Cycle\end{tabular} & \cmarkgreen & \begin{tabular}[c]{@{}c@{}}\cmarkgreen / \xmarkred\\ \small(only in $\RC$)\end{tabular} \\ \midrule
$\historyVar_5$ & 32768 & 100 & PostgreSQL & TPC-C & Future Read & \cmarkgreen & \cmarkgreen \\ \midrule
$\historyVar_6$ & 50000 & 30 & PostgreSQL & TPC-C & Future Read & \cmarkgreen & \cmarkgreen \\ \midrule
$\historyVar_7$ & 50000 & 40 & PostgreSQL & TPC-C & Future Read & \cmarkgreen & \cmarkgreen \\ \midrule
$\historyVar_8$ & 1048576 & 100 & PostgreSQL & TPC-C & Causality Cycle & \cmarkgreen & \xmarkred \\ \bottomrule
\end{tabularx}

\end{table}

\section{Related Work}\label{SEC:RELATED_WORK}

The formalization of database isolation has been a subject of continuous work following various approaches, such as axiomatically via conflict graphs and variants thereof~\cite{Terry1994a,Berenson1995,Adya2000} and operational semantics~\cite{Crooks2017}.
$\ToolName$ follows an axiomatic style using a visibility relation, initially developed in \cite{Burckhardt2014,Cerone2015}, and used by many current weak-isolation testers~\cite{Biswas2019,Liu2024a}.

The polynomial complexity of weak isolation levels admits a unifying view, as shown in~\cite{Biswas2019}. Intuitively, this stems from the fact that $\co$ appears only in one of the edges for each isolation level in \cref{fig:isolation-levels}.
This can serve as a first criterion for estimating whether a new isolation level admits polynomial-time testing.
Plume~\cite{Liu2024a} splits the problem of checking consistency into showing the absence of a number of Transactional Anomalous Patterns (TAPs), each catching a certain kind of a consistency violation that (typically) involves 3 transactions and relations between them.
The fine-grained complexity of each weak isolation level is subject to further insights specific to that level.
$\ToolName$ achieves a significant improvement in theoretical complexity and practical performance by avoiding an exhaustive search over all TAPs.

Black-box testing techniques have also been developed for strong isolation levels, most notably for Serializability~\cite{Tan2020,Geng2024} and Snapshot Isolation~\cite{Zhang2023a,Huang2023b}.
Since testing for strong isolation is NP-complete \cite{Biswas2019,Papadimitriou1979a}, these testers mostly rely on SAT/SMT solving, though more efficient algorithms exist when parameterized by the number of sessions or the communication topology~\cite{Biswas2019}.

Analogous consistency testing problems arise frequently in the context of shared-memory concurrent programs, where isolation levels give their place for memory models~\cite{Furbach2015}.
The landmark work of \cite{Gibbons1997} shows that the problem is NP-complete for Sequential Consistency, via a reduction from the Serializability isolation level~\cite{Papadimitriou1979a}.
Similar results are known for weaker memory models, such as x86-TSO, which are still relatively strong~\cite{Furbach2015}.
Nevertheless, parameterization by the number of threads and the communication topology is also known to yield polynomial-time algorithms~\cite{Gibbons1994,Abdulla2019b,Chalupa2018,Mathur2020,Bui2021}.

Causally-consistent memory models have also been manifested in shared memory, perhaps most prominently in the C/C++ memory model~\cite{Baty2011}.
Their weak semantics were shown to allow for efficient, polynomial time consistency checks~\cite{Lahav2015}, though the problem is known to become NP-complete~\cite{Bouajjani2017a}, and even notoriously difficult to parameterize~\cite{Chakraborty2024a}, when store operations do not have unique values.
On the technical level, our upper bound for $\CC$ extends a recent result for efficient consistency checks for the Strong Release-Acquire (SRA) memory model~\cite{Tunc2023} to the transactional setting.

\section{Conclusion}\label{SEC:CONCLUSION}

We have presented $\ToolName$, a highly efficient database tester for weak isolation levels.
$\ToolName$ is supported by strong theory, guaranteeing a running time of $O(\numEvents^{3/2})$, $O(\numEvents^{3/2})$, and $O(\numEvents\cdot \numSessions)$ when testing transaction histories of size $\numEvents$ and $\numSessions$ sessions, against the isolation levels Read Committed, Read Atomic, and Causal Consistency, respectively.
Moreover, we have proven that, under standard complexity-theoretic hypotheses, all weak isolation levels between Read Committed and Causal Consistency basically require at least $n^{3/2}$ time, implying that $\ToolName$ is essentially optimal.
Interesting future directions include tackling other isolation levels, possibly using saturation techniques from shared-memory concurrency~\cite{Pavlogiannis2020,Tunc2024},
as well as incorporating weak-isolation testing in a predictive analysis scheme, e.g., in the spirit of~\cite{Geng2024}.

\begin{acks}
    This work was partially supported by a research grant (VIL42117) from VILLUM FONDEN,
    and by a research grant from STIBOFONDEN.
\end{acks}

\bibliographystyle{ACM-Reference-Format}
\bibliography{bibliography}

\pagebreak

\appendix
\section{Details on \cref{SEC:ALGORITHMS}}\label{SEC:APP_ALGORITHMS}
In this section, we present details on \cref{SEC:ALGORITHMS}, including the correctness and complexity proofs of the presented algorithms.
We start with \cref{alg:internal-consistency}, which is a straightforward algorithm for checking Read Consistency (\cref{fig:internal-consistency}), and clearly runs in $O(\numEvents)$ time.
\begin{algorithm}
    \caption{Read Consistency}
    \label{alg:internal-consistency}
    
    \Fn{\FCheckInternalConsistency{$\historyVar = \tuple{\TxnSet, \so, \wr}$}}{
        \tcc{Check for thin-air reads, aborted reads, and future reads}
        \For{$\Read{x, \valVar} \in \ReadsOf{\TxnSetCommitted}$}{
            \uIf{$\Write{x, \valVar} \notin \WritesOf{\TxnSet}$}{
                \Report{thin-air read} \;
            }
            \uElseIf{$\Write{x, \valVar} \in \WritesOf{\TxnSetAborted}$}{
                \Report{aborted read} \;
            }
            \ElseIf{$\Read{x, \valVar} \relop{\po} \Write{x, \valVar}$}{
                \Report{future read} \;
            }
        }
        \tcc{Check for observe own writes and same-transaction observe latest write}
        $\mathit{lastWrites} \gets \emptyset$ \;
        \For{$\txnVar = \tuple{\OpSet, \po} \in \TxnSetCommitted$}{
            $\mathit{latestWrite} \gets \lambda x . \bot$ \;
            \For{$\opVar \in \OpSet$ in $\po$ order}{
                \Switch{$\opVar$}{
                    \uCase{$\Read{x, \valVar}$}{
                        \tcc{We assume here, that $\Read{x, \valVar}$ is not a thin-air read}
                        \uIf{$\mathit{latestWrite}[x] = \bot$ and $\txnOf{\Write{x, \valVar}} \neq \txnOf{\opVar}$}{
                            \Report{not own write} \;
                        }
                        \ElseIf{$\mathit{latestWrite}[x] \neq \Write{x, \valVar}$ and $\txnOf{\Write{x, \valVar}} = \txnOf{\opVar}$}{
                            \Report{not latest write} \tcp*{Read of stale write in own transaction}
                        }
                    }
                    \Case{$\Write{x, \valVar}$}{
                        $\mathit{latestWrite}[x] \gets \Write{x, \valVar}$ \;
                    }
                }
            }
            $\mathit{lastWrites} \gets \mathit{lastWrites} \cup \bigcup_{x} \{\mathit{latestWrite}[x]\}$
        }
        \tcc{Check for different-transaction observe latest write}
        \For{$\Read{x,\valVar} \in \ReadsOf{\TxnSetCommitted}$}{
            \If{$\txnOf{\Write{x, \valVar}} \neq \txnOf{\Read{x,\valVar}}$ and $\Write{x, \valVar} \notin \mathit{lastWrites}$}{
                \Report{not latest write} \tcp*{Read of non-final write in other transaction}
            }
        }
    }
\end{algorithm}

Next, we prove \cref{lem:saturated}, which is crucial for the correctness of our algorithms.
\lemsaturated*
\begin{proof}
For the ``only if'' direction, assume that $\historyVar$ is consistent, witnessed by the commit order $\co$, and we show that $\co'$ is acyclic (by definition, $\historyVar$ satisfies Read Consistency).
We show that $\co' \subseteq \co$ (demonstrating that $\co'$ is ``necessary''), which implies that $\co'$ is acyclic.
Since $\co'$ is minimal, any $\co'$ ordering is either contained in $\so \cup \wr$, or implied by \cref{subfig:isolation-levels-rc}. In either case the same ordering must be present in $\co$.

For the ``if'' direction, we assume that $\historyVar$ satisfies Read Consistency and $\co'$ is acyclic, and prove that $\historyVar$ is consistent by defining a suitable (total) commit order $\co$.
Here, we simply let $\co$ be any linearization of $\co'$ ($\co'$ is ``sufficient'').
Since $\co'$ is saturated and $\co' \subseteq \co$, the condition for $\IsolationLevel$ (in \cref{fig:isolation-levels}) is satisfied, meaning that $\historyVar$ is consistent.
\end{proof}

We prove correctness and complexity of \cref{alg:rc}.
\lemrcuppercorrectness*
\begin{proof}
    We prove that $\co'$ is (\ref{item:rc_upper_correctness1}) saturated and (\ref{item:rc_upper_correctness2}) minimal (\cref{def:saturated}) by the end of \FCheckRC{$\historyVar$}.
    Since we check for Read Consistency and acyclicity of $\co'$, \cref{lem:saturated} then implies correctness of the algorithm.
    Technically, parts of the algorithm only make sense if Read Consistency holds (e.g., selecting the transaction that a given read reads from).
    Thus, we shall assume that the algorithm exits, if the Read Consistency check fails on \cref{line:rc_internal_consistency}.
    For both cases, we use the invariant that at the entry of the loop on \cref{line:rc_read_loop}, $\earliestWriter$ acts, for each key $x$, as a stack of the two latest (earliest in $\po$ after $\Read{y, \valVar}$ and unique) transaction that $\txnVar_3$ has read $x$ from.
    \begin{compactenum}
        \item\label{item:rc_upper_correctness1} Clearly, \cref{line:rc_co_init} ensures that $\so \cup \wr \subseteq \co'$.
        What remains is to show that we have $\txnVar_2 \relopT{\co'} \txnVar_1$ for all transactions $\txnVar_1, \txnVar_2 \in \TxnSetCommitted$, and reads $\readVar, \readVar_x \in \ReadsOf{\TxnSetCommitted}$, where $\txnVar_1 \neq \txnVar_2$, $\txnVar_2$ writes $x$, $\txnVar_1 \relop{\wr_x} \readVar_x$, and $\txnVar_2 \relop{\wr} \readVar \relop{\po} \readVar_x$.
        Let $\txnVar_3$ be the transaction containing $\readVar$ and $\readVar_x$, and let $\readVar'$ be the $\po$-first read of $\txnVar_2$ by $\txnVar_3$.
        When $\txnVar_3$ is processed in the outer loop, the algorithm will add $\readVar' \in \firstTxnReads$.
        Hence, when $\readVar'$ is processed on \cref{line:rc_read_loop}, we enter the loop on \cref{line:rc_x_loop}.
        When $x$ is processed in this loop, an ordering $\txnVar_2 \relop{\co'} \txnVar_1'$ is added, where $\txnVar_1'$ is the next (in $\po$ after $\readVar'$) transaction that $\txnVar_3$ reads $x$ from.
        If $\txnVar_1' = \txnVar_1$, we are done.
        Otherwise, we can repeat the argument by setting $\txnVar_2 \triangleq \txnVar_1'$, which yields another transaction $\txnVar_1''$ writing to $x$ with $\txnVar_1' \relop{\co'} \txnVar_1''$.
        We will thus eventually have $\txnVar_2 \relopT{\co'} \txnVar_1$.
        \item\label{item:rc_upper_correctness2}
        This is obvious from inspecting the \textbf{if} and \textbf{for} conditions that hold on \cref{line:rc_co_update}, where $\co'$ is updated, given that the invariant on $\earliestWriter$ holds.
    \end{compactenum}
    \vspace{-1.5em}
\end{proof}

\lemrcuppercomplexity*
\begin{proof}
    \cref{alg:internal-consistency} costs $O(\numEvents)$, and the final acyclicity check on \cref{line:rc_cycle_check} can be charged to the total running time for \cref{line:rc_co_update}.
    The total time (across the entire execution) for the loop on \cref{line:rc_r_loop} will clearly be $\sum_{\txnVar_3 \in \TxnSetCommitted} |(\ReadsOf{\txnVar_3})| = O(\numEvents)$.
    Therefore, the loop on \cref{line:rc_read_loop} dominates the running time of the algorithm.
    This loop runs $O(\numEvents)$ times in total, so we focus on those iteration, where we enter the loop on \cref{line:rc_x_loop}.
    We analyze the running time by separately counting the time for those transactions $\txnVar_3$, where $|\keysRead{\txnVar_3}| \leq \sqrt{\numEvents}$ or not.

    Considering first transactions $\txnVar_3$ such that $|\keysRead{\txnVar_3}| > \sqrt{\numEvents}$, notice that there are less than $\sqrt{\numEvents}$ of these.
    Notice also that $\sum_{\txnVar_2 \relop{\wr} \txnVar_3} |\keysWritten{\txnVar_2}| = O(\numEvents)$. Hence, the total running time for the innermost loop for these transactions is $O(\numEvents^{3/2})$.

    Now consider those $\txnVar_3$ where $|\keysRead{\txnVar_3}| \leq \sqrt{\numEvents}$.
    We use the fact that the innermost loop is only entered once for every edge $\txnVar_2 \relop{\wr} \txnVar_3$, namely when $\Read{y,\valVar}$ is the first read of $\txnVar_2$ by $\txnVar_3$.
    Since $\readKeys \subseteq \keysRead{\txnVar_3}$, the total number of iterations is bounded by
    \[
        \sum_{\txnVar_3 \in \TxnSetCommitted} \sum_{\txnVar_2 \relop{\wr} \txnVar_3} |\keysRead{\txnVar_3}| \leq \sum_{\txnVar_3 \in \TxnSetCommitted} |\keysRead{\txnVar_3}|^2
    \]
    This sum is maximized when each $|\keysRead{\txnVar_3}|$ is as big as possible, i.e., $|\keysRead{\txnVar_3}| = \sqrt{\numEvents}$.
    But this also implies that $|\TxnSetCommitted| \leq \sqrt{\numEvents}$, hence the total becomes
    \[
        \sum_{\txnVar_3 \in \TxnSetCommitted} |\keysRead{\txnVar_3}|^2 \leq \sum_{\txnVar_3 \in \TxnSetCommitted} \numEvents = \numEvents^{3/2}
    \]

    In conclusion, both categories of transactions take $O(\numEvents^{3/2})$ time in total.
\end{proof}

We prove the correctness and complexity of \cref{alg:ra}.
\lemrauppercorrectness*
\begin{proof}
    We prove that $\co'$ is (\ref{item:ra_upper_correctness1}) saturated and (\ref{item:ra_upper_correctness2}) minimal (\cref{def:saturated}) by the end of \FCheckRA{$\historyVar$}.
    Since we check for Read Consistency and acyclicity of $\co'$, \cref{lem:saturated} then implies correctness of the algorithm.
    After checking Read Consistency, the algorithms checks for the \emph{repeatable reads} property, and we assume termination if this does not hold.
    In short, this property states that a transaction cannot read a key from two different transactions.
    The procedure for checking this (\FCheckRepeatableReads) is straight forward, and we assume this property from this point.
    \begin{compactenum}
        \item\label{item:ra_upper_correctness1} Clearly, \cref{line:ra_co_def} ensures that $\so \cup \wr \subseteq \co'$.
        What remains is to show that we have $\txnVar_2 \relopT{\co'} \txnVar_1$ for all transactions $\txnVar_1, \txnVar_2, \txnVar_3 \in \TxnSetCommitted$, where $\txnVar_1 \neq \txnVar_2$, $\txnVar_2$ writes $x$, $\txnVar_1 \relop{\wr_x} \txnVar_3$, and $\txnVar_2 \relop{\so\,\cup\,\wr} \txnVar_3$.
        Consider first if $\txnVar_2 \relop{\so} \txnVar_3$.
        The algorithm will eventually iterate the read $\txnVar_1 \relop{\wr_x} \txnVar_3$ on \cref{line:ra_t1_loop}.
        At this point, an ordering $\txnVar_2' \relop{\co'} \txnVar_1$ is added, where $\txnVar_2'$ is the last transaction writing $x$ $\so$-before $\txnVar_3$.
        We have $\txnVar_2 \relopR{\so} \txnVar_2'$, and thus also $\txnVar_2 \relopT{\co'} \txnVar_1$.
        Next, consider if $\txnVar_2 \relop{\wr} \txnVar_3$, which will eventually be iterated on \cref{line:ra_t2_loop}.
        Due to the uniqueness ensured by repeatable reads, $\txnVar_1$ will be chosen on \cref{line:ra_t1_def}, and $\txnVar_2 \relop{\co'} \txnVar_1$ is added directly.
        \item\label{item:ra_upper_correctness2} This is obvious from inspecting the \textbf{if} and \textbf{for} conditions that hold on \cref{line:ra_co_update_so} and \cref{line:ra_co_update_wr}, where $\co'$ is updated.
    \end{compactenum}
    \vspace{-1.5em}
\end{proof}

\lemrauppercomplexity*
\begin{proof}
    \cref{alg:internal-consistency} and \FCheckRepeatableReads costs $O(\numEvents)$, and the final acyclicity check on \cref{line:ra_acyclicity_check} can be charged to the total running time for \cref{line:ra_co_update_so} and \cref{line:ra_co_update_wr}.
    The total time (across the entire execution) for the loop on \cref{line:ra_t1_loop} will clearly be $\sum_{\txnVar_3 \in \TxnSetCommitted} |(\ReadsOf{\txnVar_3})| = O(\numEvents)$.
    Therefore, the innermost loop on \cref{line:ra_x_loop} dominates the running time of the algorithm.
    We analyze the running time by separately counting the time for those transactions $\txnVar_3$, where $|\keysRead{\txnVar_3}| \leq \sqrt{\numEvents}$ or not.

    Considering first transactions $\txnVar_3$ such that $|\keysRead{\txnVar_3}| > \sqrt{\numEvents}$, notice that there are less than $\sqrt{\numEvents}$ of these.
    Notice also that $\sum_{\txnVar_2 \relop{\wr} \txnVar_3} |\keysWritten{\txnVar_2}| = O(\numEvents)$. Hence, the total running time for the innermost loop for these transactions is $O(\numEvents^{3/2})$.

    Now consider those $\txnVar_3$ where $|\keysRead{\txnVar_3}| \leq \sqrt{\numEvents}$.
    The total number of iterations in this case is bounded by
    \[
        \sum_{\txnVar_3 \in \TxnSetCommitted} \sum_{\txnVar_2 \relop{\wr} \txnVar_3} |\keysRead{\txnVar_3}| \leq \sum_{\txnVar_3 \in \TxnSetCommitted} |\keysRead{\txnVar_3}|^2
    \]
    This sum is maximized when each $|\keysRead{\txnVar_3}|$ is as big as possible, i.e., $|\keysRead{\txnVar_3}| = \sqrt{\numEvents}$.
    But this also implies that $|\TxnSetCommitted| = \sqrt{\numEvents}$, hence the total becomes
    \[
        \sum_{\txnVar_3 \in \TxnSetCommitted} |\keysRead{\txnVar_3}|^2 \leq \sum_{\txnVar_3 \in \TxnSetCommitted} \numEvents = \numEvents^{3/2}
    \]

    In conclusion, both categories of transactions take $O(\numEvents^{3/2})$ time in total.
\end{proof}

Finally, we prove the correctness and complexity of \cref{alg:cc}.
\lemcccorrectness*
\begin{proof}
    We prove that $\co'$ is (\ref{item:ra_upper_correctness1}) saturated and (\ref{item:ra_upper_correctness2}) minimal (\cref{def:saturated}) by the end of \FCheckCC{$\historyVar$}.
    Since we check for Read Consistency and acyclicity of $\co'$, \cref{lem:saturated} then implies correctness of the algorithm.
    The computation of $\HB$ (\FComputeHB) is standard, so we skip proving its correctness.
    We use an invariant that, after processing $\txnVar_3$, $\lastWrite{\sessionVar'}$ contains, for each key $x$, the $\so$-last transaction $\txnVar_2'$ of $\sessionVar'$ such that $\txnVar_2' \relopT{\so\,\cup\,\wr} \txnVar_3$.
    An important property is that any following transaction $\txnVar_3'$ also has $\txnVar_2' \relopT{\so\,\cup\,\wr} \txnVar_3'$ for each of these $\txnVar_2'$.
    \begin{compactenum}
        \item\label{item:ra_upper_correctness1}  Clearly, \cref{line:algo_cc_initco} ensures that $\so \cup \wr \subseteq \co'$.
        What remains is to show that we have $\txnVar_2 \relopT{\co'} \txnVar_1$ for all transactions $\txnVar_1, \txnVar_2, \txnVar_3 \in \TxnSetCommitted$, where $\txnVar_1 \neq \txnVar_2$, $\txnVar_2$ writes $x$, $\txnVar_1 \relop{\wr_x} \txnVar_3$, and $\txnVar_2 \relopT{\so\,\cup\,\wr} \txnVar_3$.
        Consider the iteration of the loop on \cref{line:algo_cc_iterate_wr} that processes $\txnVar_1 \relop{\wr_x} \txnVar_3$.
        When iteration the session $\sessionVar'$ of $\txnVar_2$, we will then have $\txnVar_2 \relopR{\so} \txnVar_2' = \lastWrite{\sessionVar'}[x]$.
        If $\txnVar_2' = \txnVar_1$, the desired ordering is implied by $\so$, and otherwise we add $\txnVar_2' \relop{\co'} \txnVar_1$.
        In either case $\txnVar_2 \relopT{\co'} \txnVar_1$.
        \item\label{item:ra_upper_correctness2} This is obvious from inspecting the \textbf{if} and \textbf{for} conditions that hold on \cref{line:algo_cc_saturate}, where $\co'$ is updated.
    \end{compactenum}
    \vspace{-1.5em}
\end{proof}

\lemcccomplexity*
\begin{proof}
We have already argued that checking for Read Consistency in \cref{line:algo_cc_read_consistency} runs in $O(\numEvents)$ time.
The computation of $\HB$ by $\FComputeHB(\historyVar)$ runs in $O(\numEvents\cdot \numSessions)$ time, dominated by $O(\numEvents)$ join operations on Vector Clocks (one for each read event in $\historyVar$), each taking $O(\numSessions)$ time.
The main algorithm iterates over $O(\numEvents)$ orderings $\txnVar_1\relop{\wr}\txnVar_3$ in \cref{line:algo_cc_iterate_wr}, once for each read event in $\historyVar$.
For each such edge, it performs $O(\numSessions)$ time on average, since $\lastWrite{x}$ scans the writer list $\Writes{\sessionVar'}[x]$ in one pass.
Thus the total time is $O(\numEvents\cdot \numSessions)$, as desired.
\end{proof}
\section{Details on \cref{SEC:LOWER_BOUNDS}}\label{SEC:APP_LOWER_BOUNDS}

In this section we present the detailed proofs of \cref{lem:range_lower_correctness}, \cref{lem:ra_lower_correctness} and \cref{lem:rc_lower_correctness}, as well as \cref{thm:ra_one_session_linear_time}.

\lemrangelowercorrectness*
\begin{proof}
We prove each item separately.

\begin{compactenum}
\item We prove the contrapositive.
Assume that $\historyVar = \tuple{\TxnSet, \so, \wr}$ violates $\CC$, and we will show that $G = \tuple{V, E}$ contains a triangle.
Each condition of Read Consistency hold trivially, so there must be no $\co$ respecting $\so \cup \wr$ that satisfies the $\CC$ axiom (\cref{subfig:isolation-levels-cc}).
Let $\co$ be any commit order that respects $\so \cup \wr$.
There must then be $x \in \KeySet,\txnVar_1,\txnVar_2,\txnVar_3 \in \TxnSetCommitted$ such that $\txnVar_1 \neq \txnVar_2$, $\txnVar_1 \relop{\wr_x} \txnVar_3$, $\txnVar_2$ writes $x$, $\txnVar_2 \relopT{\so\,\cup\,\wr} \txnVar_3$, and $\txnVar_1 \relop{\co} \txnVar_2$.
Since $\txnVar_1 \relop{\wr_x} \txnVar_3$, we must have that $\txnVar_3 = \txnVar_c^\R$ for some $c \in V$.
Further, since $\txnVar_2 \neq \txnVar_1$ and $\txnVar_2$ writes $x$, it must be that $x = x_a$, $\txnVar_1 = \txnVar_a^\W$, and $\txnVar_2 = \txnVar_b^\W$ for some $a,b \in V$ with $a \neq b$, because the key $x_a^c$ is only written in $\txnVar_a^\W$.
Finally, we have $\txnVar_2 \relop{\wr} \txnVar_3$, since $\so=\emptyset$ and $\wr^+ = \wr$.
The three facts 
(i)~$\txnVar_a^\W \relop{\wr} \txnVar_c^\R$, 
(ii)~$\txnVar_b^\W$ writing $x_a$, and 
(iii)~$\txnVar_b^\W \relop{\wr} \txnVar_c^\R$ imply that
(i)~$\tuple{a,c} \in E$, 
(ii)~$\tuple{a,b} \in E$, and
(iii)~$\tuple{b,c} \in E$, respectively.
This constitutes a triangle in $G$.

\item  We again prove the contrapositive.
Assume that $G$ forms a triangle between nodes $a,b,c\in V$, and we will argue that $\historyVar$ is inconsistent with $\RC$.
Since $\tuple{a,c}, \tuple{b,c}\in E$, we have $\txnVar_b^\W \relop{\wr} \Read{x_b^c, b} \relop{\po} \Read{x_a, a}$, where $\Read{x_b^c, b}, \Read{x_a, a}$ are operations of $\txnVar_c^\R$.
We also have $\txnVar_a^\W \relop{\wr} \Read{x_a, a}$, and since $\tuple{a,b} \in E$, $\txnVar_b^{\W}$ writes $x_a$.
Hence, any valid commit order $\co$, must have $\txnVar_b^\W \relop{\co} \txnVar_a^\W$.
Using a symmetric argument by exchanging $b$ and $a$, we can argue that $\co$ must order $\txnVar_a^\W \relop{\co} \txnVar_b^\W$.
Therefore, no valid commit order can exist, and $\historyVar$ must be inconsistent with $\RC$.
\end{compactenum}
\vspace{-1.5em}
\end{proof}

\lemralowercorrectness*
\begin{proof}
We prove the contrapositives of the two implications: (\ref{item:ra_lower1}) if $\historyVar$ is \emph{not} consistent with $\RA$, $G$ has a triangle, and (\ref{item:ra_lower2}) if $G$ has a triangle, $\historyVar$ is \emph{not} consistent with $\RA$.
\begin{compactenum}
    \item\label{item:ra_lower1} Assume that $\historyVar = \tuple{\TxnSet, \so, \wr}$ violates $\RA$, and we show that $G = \tuple{V, E}$ contains a triangle.
    Read Consistency holds trivially, and there is no $\co$ respecting $\so \cup \wr$ that satisfy the $\RA$ axiom.
    Let $\co$ be any such commit order.
    There must be $x \in \KeySet,\txnVar_1,\txnVar_2,\txnVar_3 \in \TxnSetCommitted$ such that $\txnVar_1 \neq \txnVar_2$, $\txnVar_1 \relop{\wr_x} \txnVar_3$, $\txnVar_2$ writes $x$, $\txnVar_2 \relop{\so\,\cup\,\wr} \txnVar_3$, and $\txnVar_1 \relop{\co} \txnVar_2$.
    For the same reasons as in the proof of \cref{lem:range_lower_correctness}, we must have $x = x_a$, $\txnVar_1 = \txnVar_a^\W$, $\txnVar_2 = \txnVar_b^\W$, and $\txnVar_3 = \txnVar_c^\R$ for some nodes $a,b,c\in V$.
    It also still holds that $\txnVar_2 \relop{\wr} \txnVar_3$, since $\txnVar_2$ is on the $\sessionVar_\W$ session and $\txnVar_3$ is on the $\sessionVar_\R$ session.
	The three facts
	(i)~$\txnVar_a^\W \relop{\wr} \txnVar_c^\R$,
	(ii)~$\txnVar_b^\W$ writing $x_a$, and
	(iii)~$\txnVar_b^\W \relop{\wr} \txnVar_c^\R$,
	imply the existence of a triangle $\tuple{a,b,c}$.
    \item\label{item:ra_lower2} Assume that $G = \tuple{V, E}$ has a triangle between the nodes $a,b,c \in V$.
    We show that $\historyVar = \tuple{\TxnSet, \so, \wr}$ violates $\RA$. Since $\tuple{a,c}, \tuple{b,c} \in E$, we have $\txnVar_a^\W \relop{\wr_{x_a}} \txnVar_c^\R$ and $\txnVar_b^\W \relop{\wr} \txnVar_c^\R$. 
	Since $\tuple{b,a} \in E$, we have that $\txnVar_b^\W$ writes $x_a$, hence any valid $\co$ must have $\txnVar_b^\W \relop{\co} \txnVar_a^\W$. 
	Symmetrically, we can derive $\txnVar_a^\W \relop{\co} \txnVar_b^\W$, which means that no valid $\co$ exists.
\end{compactenum}
\vspace{-1.5em}
\end{proof}

\thmraonesessionlineartime*
\begin{proof}
First, Read Consistency can be checked in $O(\numEvents)$ time, as demonstrated by \cref{alg:internal-consistency}.
Similarly, the acyclicity of $\so \cup \wr$ requires $O(\numEvents)$ time.
Notice that, since $\co$ has to respect $\so \cup \wr$, we must simply have $\co = \so$.
It thus remains to check the $\RA$ axiom (\cref{subfig:isolation-levels-ra}) for this $\co$.
We can rephrase this task as checking, for each read $\txnVar_1 \relop{\wr_x} \txnVar_3$, whether there is $\txnVar_2$ writing $x$ such that $\txnVar_1 \relop{\co} \txnVar_2 \relop{\co} \txnVar_3$.
If such $\txnVar_2$ exists, the $\RA$ axiom says that $\txnVar_2 \relop{\co} \txnVar_1$ should hold, a contradiction.
This check can be done by scanning all transactions in $\co$ order and maintaining the latest write to each key seen.
\end{proof}

\lemrclowercorrectness*
\begin{proof}
We again prove the contrapositives of the two implications: (\ref{item:rc_lower1}) if $\historyVar$ is \emph{not} consistent with $\RC$, $G$ has a triangle, and (\ref{item:rc_lower2}) if $G$ has a triangle, $\historyVar$ is \emph{not} consistent with $\RC$.
\begin{compactenum}
    \item\label{item:rc_lower1} Assume that $\historyVar = \tuple{\TxnSet, \so, \wr}$ violates $\RC$, and we show that $G = \tuple{V, E}$ contains a triangle.
    Each condition of Read Consistency holds trivially, so we turn our attention to $\co$.
    We let $\co = \so$, as they must agree.
    We then get that there are $x \in \KeySet, \txnVar_1, \txnVar_2 \in \TxnSetCommitted, \readVar, \readVar_x \in \ReadsOf{\TxnSetCommitted}$ such that $\txnVar_1 \neq \txnVar_2$, $\txnVar_1 \relop{\wr_x} \readVar_x$, $\txnVar_2$ writes $x$, $\txnVar_2 \relop{\wr} \readVar \relop{\po} \readVar_x$, and $\txnVar_1 \relop{\co} \txnVar_2$.
    If we let $\txnVar_3$ be the transaction that contains $\readVar$ and $\readVar_x$, we then have $\txnVar_1 \relop{\wr} \txnVar_3$ and $\txnVar_2 \relop{\wr} \txnVar_3$.
    By the proof of \cref{lem:range_lower_correctness} (\ref{item:range_lower1}), these conditions, along with $\txnVar_2$ writing $x$, are sufficient to show that $G$ has a cycle.
    \item\label{item:rc_lower2} The proof of \cref{lem:range_lower_correctness} (\ref{item:range_lower2}) does not rely on $\so$, hence it holds here as well.
\end{compactenum}
\vspace{-1.5em}
\end{proof}

\end{document}